\documentclass[12pt,color]{article}
\usepackage{amsmath}
\usepackage{amssymb}
\usepackage{color}
\usepackage{xcolor,cancel}
\usepackage[normalem]{ulem}
\numberwithin{equation}{section} 

\definecolor{forestgreen}{rgb}{0.13,0.54,0.13}

\newcommand{\comments}[1]{}

\newcommand{\rme}{{\rm e}}
\newcommand{\rmd}{{\rm d}}
\newcommand{\rmi}{{\rm i}}
\newcommand{\rmr}{{\rm r}}
\newcommand{\ud}[1]{\hspace{-0em}\mathrm{d}{#1}\;}
\newcommand{\udd}[2]{\hspace{-0em}\mathrm{d}{#1}\,\mathrm{d}{#2}\;}

\newcommand{\uD}{\mathcal{D}}
\newcommand{\eff}{\rm{eff}}

\begin{document}

\title{Dynamical symmetries of Markov processes with multiplicative white noise}
\author{
Camille Aron$^{1,2}$, 
Daniel G.\ Barci$^3$,
Leticia F. Cugliandolo$^4$, \\
Zochil Gonz\'alez Arenas$^3$, 
and Gustavo S. Lozano$^{4,5}$
\bigskip
\\
$^1${\small Department of Electrical Engineering, Princeton University, Princeton, NJ 08544, USA}\\
$^2${\small Instituut voor Theoretische Fysica, KU Leuven, Belgium}
\\
$^3${\small Departamento de F{\'\i}sica Te\'orica, Universidade do Estado do Rio de Janeiro,}
\\ {\small Rua S\~ao Francisco Xavier 524, 20550-013, Rio de Janeiro, RJ, Brazil}
\\
$^4${\small Sorbonne Universit\'es, Universit\'e Pierre et Marie Curie, UMR 7589}
\\
{\small  Laboratoire de Physique Th\'eorique et Hautes Energies, 
Paris, France}
\\
$^5${\small Departamento de F\'{\i}sica, FCEYN Universidad de Buenos Aires \& IFIBA CONICET,} \\
{\small Pabell\'on 1 Ciudad Universitaria, 1428 Buenos Aires, Argentina}
}

\date{}

\maketitle

\newpage

\begin{abstract}
We analyse various properties of stochastic Markov processes with multiplicative white noise. 
We take a single-variable problem as a simple example, and we later extend the analysis to the Landau-Lifshitz-Gilbert equation for the stochastic dynamics of a magnetic moment. In particular, we focus on the non-equilibrium transfer of angular momentum to the magnetization from a spin-polarised current of electrons, a technique which is widely used in the context of spintronics to manipulate magnetic moments.
We unveil two hidden dynamical symmetries of the generating functionals of these Markovian multiplicative white-noise processes. One symmetry only holds in equilibrium and we use it to prove generic relations such as the fluctuation-dissipation theorems. Out of equilibrium, we take profit of the symmetry-breaking terms to prove fluctuation theorems. The other symmetry yields strong dynamical relations between correlation and response functions which can notably simplify the numerical analysis of these problems.
 Our construction allows us to clarify some misconceptions on multiplicative white-noise stochastic processes that can be found in the literature. In particular, we show that a first-order differential equation with multiplicative white noise can be transformed into an additive-noise equation, but that the latter keeps a non-trivial memory of the discretisation prescription used to define the former.
\end{abstract}














\newpage

\tableofcontents 

\vspace{1cm}

\section{Introduction}

Stochastic Markov processes in which the noise acts multiplicatively on a function of the variable of interest are manifold.
In physics one finds the diffusion of a colloidal particle close to a wall, in chemistry
one counts autocatalytic chemical reactions in which the production of a molecule is enhanced by the presence of the same molecules already produced, in economy the Black and Scholes model provides a theory of option pricing.
A detailed discussion of the dynamics of single variable and extended systems with multiplicative noise can be found in Ref.~\cite{Sagues07}.

{In order to make sense, any Markovian stochastic equation with multiplicative noise, \textit{e.g.} an overdamped Langevin equation with state-dependent diffusion coefficient,  must be given a discretization prescription.} For these, the 
associated Fokker-Planck equation governing the time evolution of
the probability distribution function of the stochastic variable(s)
depends, in general, on the discretisation prescription 
parameter, say $\alpha$, and on the function that multiplies the noise, say $g$. Noteworthy, its stationary solution 
also depends on $\alpha$ and $g$~\cite{Strato,gardiner,vanKampen,Oksendhal}.

In the presence of a multiplicative noise, Stratonovich noticed that the qualitative behaviour of the stationary probability distribution can change as a function of the noise strength~\cite{Strato}, and thus deviate from the usual Gibbs-Boltzmann distribution which would only involve the potential responsible for the deterministic forces acting, say, on the particle. The state-dependence of the function $g$ can have far reaching consequences.
{
 For instance, the stationary probability distribution function may develop new extrema that are not the ones set by the deterministic forces.
Similar effect on many-body systems can alter the number or the nature of the extrema of the free energy governing the dynamics of an order parameter, and therefore induce so-called noise-induced phase transitions~\cite{Sagues07}. For instance, models without symmetry breaking potentials can thus exhibit coarsening phenomena.}

However, one can adopt a different point of view from the one above. The stochastic differential equation  can be modified so that the approach 
to the usual Gibbs-Boltzmann distribution is ensured for any discretisation prescription parameter $\alpha$ and multiplicative function $g$.
This is achieved by adding a drift term to the stochastic equation [given in Eq.~(\ref{eq:equil-choice-f}) 
and necessary even when the common Stratonovich mid-point prescription is used]. With this addition, the $\alpha$ dependence 
disappears from the Fokker-Planck equation (and the physics in general). {Although the dynamics still depend on $g$, 
one can show that they converge to a $g$-independent stationary solution which is now the desired Gibbs-Boltzmann measure.}

We present a concise summary of these known, though perhaps not sufficiently assimilated, issues in the two first subsections of 
Sec.~\ref{sec:one-dim}, 
that are supplemented by material  in Apps.~\ref{subsec:chain rule}, \ref{app:mult-addit} and \ref{subsec:stat-Langevin}, 
in the framework of a stochastic differential equation on a single variable.

{
It is sometimes found in the literature that one-dimensional Markov processes with multiplicative noise can 
be mapped to overdamped Langevin equations with an additive noise}, and that once in this new framework all discretisation subtleties can be ignored. 
This statement is, however, wrong as the new additive-noise equation depends explicitly on the $\alpha$-prescription used to define 
the original multiplicative-noise equation. 
The reason is that the chain rule for the time derivative of a function of the stochastic variable has to be used
in the transformation between multiplicative and additive-noise equations, and this chain rule involves $\alpha$ and $g$. 
Accordingly, the Fokker-Planck equation associated to the resulting
additive-noise Langevin equation and its asymptotic solution depend on $\alpha$ and $g$.
This can be cured by adding a drift term to the additive-noise Langevin equation 
that is completely equivalent to the one to be used in the multiplicative-noise formalism. We discuss these facts in 
Apps.~\ref{app:mult-addit} and \ref{subsec:stat-Langevin}.

{In Sec.~\ref{subsec:path-int} we recall the path-integral generating-functional formalism for stochastic Langevin processes with multiplicative white noise~\cite{Leschke77,Langouche79,Langouche81,Tirapegui-book, Arnold2000,Lubensky2007}}. We explain the apparent differences with the path integrals used in Ref.~\cite{Jarzynski08}, and why we do not agree with the claims in Refs.~\cite{Ao14a,Ao14b}. 

The functional formulation of stochastic processes is very well-suited to prove model-inde\-pen\-dent properties of generic physical observables.  {
Although the derivation of fluctuation theorems for white-noise Markov processes has been addressed on general grounds \textit{via} equation-of-motion formalisms~\cite{ChetriteGawedzki,Jarzynski08, Chetrite2011}, we are still lacking a generic path-integral formulation addressing the case of multiplicative noise.}
In this paper, we will use a {model-independent field transformation in} the path-integral formulation to show that the equilibrium fluctuation-dissipation theorems and the out-of-equilibrium fluctuation relations~\cite{Jarzynski97,Kurchan98,Lebowitz99,Evans02, Ritort03, Maes03,Park04, Bustamante05,seifert2008,Zamponi} hold for multiplicative white-noise Markov processes.
{
Contrary to previous works for which the steady states were governed by non-equilibrium potentials, see \textit{e.g.}~\cite{Jarzynski08},~\cite{Ao14a,Ao14b} and~\cite{Zochil10,arenas2012}, here the approach to a Gibbs-Boltzmann equilibrium is ensured by the presence of a drift term, see also~\cite{Chetrite}.}

We will also present another transformation which leaves the action (and the functional measure) invariant and which can be used to derive the Schwing\-er-Dyson equations governing the coupled dynamics of correlations and linear responses.

For simplicity, we will present detailed derivations in the framework of a single-variable stochastic equation. The generalisation to higher dimensional problems, and field theories, should then be clear. At the end of the paper, we will apply our results to the dynamics of a magnetic moment governed by the stochastic Landau-Lifshitz-Gilbert equation~\cite{Langevin-Coffey,Bertotti-etal} taking advantage of the path-integral formalism developed in Ref.~\cite{Aron14}. 

Recapping, the paper is organised as follows. In Sec.~\ref{sec:one-dim}, we recall the main features of the Langevin, Fokker-Planck and path-integral formulations of stochastic Markov processes with multiplicative white noise in the framework of single-variable problems.
We re-derive Crooks relation~\cite{Crooks00} between 
path probabilities for forward and backward stochastic processes, now extended to take into account non-trivial issues due to the discretisation of Markov stochastic processes with multiplicative white noise.  We prove equilibrium and out-of-equilibrium fluctuation theorems.
Complements to these sections are given in the Appendices. Section~\ref{sec:LLG} is devoted to the application of these ideas to the Landau-Lifshitz-Gilbert stochastic equation. Finally, in Sec.~\ref{sec:conclusions} we open some ways for future research.

\section{Single-variable stochastic Markov processes}
\label{sec:one-dim}

In this Section, we define the single-variable problem that we use as a framework to recall a number of important features that, sometimes, appear in confusing terms in the literature.
We also discuss a time-reversal symmetry of the equilibrium generating functional and we use it to derive 
equilibrium relations such as the fluctuation-dissipation theorem. Out of equilibrium, this symmetry is broken and we use the resulting symmetry-breaking terms to derive fluctuation relations.
We then discuss another symmetry of the generating functional, valid in and out of equilibrium, that is useful to derive the Schwinger-Dyson equations for correlations and linear responses.

\subsection{The Langevin equation}

Let us consider a real variable $x$, the dynamics of which is governed by 
the following Langevin equation
\begin{equation}
\rmd_t x(t) = f(x) + g(x) \xi(t)   \label{eq:x-eom}
\end{equation}
where $\xi(t)$ is a Gaussian white noise
with zero mean $\langle \xi(t) \rangle = 0$ and variance 
\begin{align}
\langle \xi(t) \xi(t') \rangle = 2 D \delta(t-t') \mbox{ with } D=k_B T \;.
\end{align}
The noise is said to be \emph{multiplicative} because it acts multiplicatively on $g(x)$, a function of the stochastic variable.

This stochastic differential equation makes sense only when complemented with a discretisation prescription to define at which point $g(x)$ should be evaluated. This is relevant since each pulse $\xi(t)$ yields a discontinuity in $x$ and therefore the value of $x$ at which $g(x)$ is evaluated (and hence the size of the discontinuity) is \textit{a priori} not well defined.
Without restricting the generality of the foregoing, we work with the generic $\alpha$-prescription~\cite{gardiner,Langouche78} which corresponds, in discrete time, to
\begin{equation}
x_{n+1}-x_n = f(\overline x_n) \rmd t+ g(\overline x_n) {\rm d}W_n
\end{equation}
with ${\rm d}W_n \equiv \xi_n {\rm d}t$, $\langle \rmd W_n\rangle =0$ and $\langle \rmd W_n \rmd W_m \rangle = 2D \delta_{nm} \rmd t$ for the statistics of the noise, 
and
\begin{equation}
\overline x_n = \alpha x_{n+1} + (1-\alpha) x_n 
\; , 
\end{equation}
with $\alpha$ a real parameter $0 \leq \alpha \leq 1$. Note that the discretisation used in the argument of 
$f$ is irrelevant in the continuous-time limit.
Equation~(\ref{eq:x-eom}) can be taken into a form in which the noise appears additively but in which the dependence on the discretisation parameter $\alpha$ is still present in the new equation, see Apps.~\ref{app:mult-addit} and \ref{subsec:stat-Langevin}.

{The chain rule for the time derivative of a function $F$ of the variable $x$ depends on the stochastic equation governing the time evolution of $x$, see Refs.~\cite{gardiner,vanKampen,Oksendhal} for It\^o ($\alpha=0$) and Stratonovich ($\alpha=1/2$) prescriptions and App.~\ref{subsec:chain rule} for a generic $\alpha$-prescription. In this case, it reads}
\begin{align} \label{eq:x-chain}
\rmd_t F(x)= \rmd_t x \ \partial_x F(x) + (1-2\alpha) D  g^2(x)  \partial^2_x F(x)
\end{align}
where ${\rm d}_t x \equiv {\rm d}x/{\rm d}t$. 
The usual chain rule of conventional calculus is recovered only in the case of the Stratonovich mid-point prescription, $\alpha=1/2$.
Note that the chain rule is independent of the `force' $f(x)$. In particular, it does not depend on the addition (or not) of a drift term to the Langevin equation such as discussed around Eq.~(\ref{eq:x-eom-drifted}) below.

\subsection{The Fokker-Planck equation}
\label{subsec:FP}

The Fokker-Planck equation corresponding to the Langevin equation (\ref{eq:x-eom}) in a generic
$\alpha$-prescription reads~\cite{gardiner,vanKampen,Langouche79,arenas2012}
\begin{eqnarray}
\partial_t P(x,t)
 = 
- \partial_x[ (f(x) + 2D\alpha g(x) g'(x)) P(x,t) ] 
+ D \, \partial_x^2 [ g^2(x) P(x,t) ]
\; .
\label{eq:FP-without-drift}
\end{eqnarray}
(Note the difference between this equation and the one  used in Ref.~\cite{Jarzynski08}.)
Once supplemented by an initial condition $P_{\rm i}(x) = P(x,0)$, this equation describes the deterministic evolution of the probability density $P(x,t)$ of finding $x$ at 
time $t$. It can be written in the form of a continuity equation $\partial_t P + \partial_x J = 0$.  Its stationary solution with vanishing current, $J=0$, is
\begin{equation}
P_{\rm st}(x) = Z^{-1} \ [g(x)]^{2(\alpha-1)} \ \rme^{\frac{1}{D} \int^x \frac{f(x')}{g^2(x')}}
\label{eq:stat-sol-no-drift}
\end{equation}
where $\int^x$ represents the indefinite integral over $x'$ and $Z$ is a normalisation constant~\cite{gardiner,vanKampen,arenas2012}. 
The  approach to this asymptotic form can be proven with the construction of an ${\cal H}$-function as in Ref.~\cite{DoiEdwards}, or with the mapping of the Fokker-Planck operator into a Schr\"odinger operator and the analysis of its eigenvalue problem~\cite{Parisi}.
Clearly, the fact that $P_{\rm st}$ depends on $\alpha$ and $g$ shows that these can have highly non-trivial consequences on the transient dynamics as well as the asymptotic stationary properties of the system~\cite{Sagues07,Strato,Horsthemke84}.

However, if we allow ourselves to consider the special `drift force'~\cite{Klimontovich1990}
\begin{equation}
f(x) = - g^2(x) V'(x) + 2D (1-\alpha)  g(x) g'(x) \; , 
\label{eq:equil-choice-f}
\end{equation}
with the short-hand notation $V'\equiv\partial_x V$ and $g'\equiv\partial_x g$, 
the Fokker-Planck  equation loses any dependence on $\alpha$, 
\begin{eqnarray}
&&
\partial_t P(x,t)
=
 \partial_x \{ g^2(x) [ V'(x)  P(x,t) + D \, \partial_x P(x,t) ]
\}
\;.
\label{eq:FP-correct}
\end{eqnarray}
Importantly, since physical observables are computed using $P(x,t)$, this implies that the physics of Eq.~(\ref{eq:x-eom}) with the `drift force' in Eq.~(\ref{eq:equil-choice-f})
does not depend on the prescription parameter $\alpha$. 
This can also be proven using the BRST symmetry of the generating functional~\cite{Zochil10} or with a perturbative analysis~\cite{Honkonen}. 
Moreover, the asymptotic solution of the Fokker-Planck equation in Eq.~(\ref{eq:FP-correct}) simply reads
\begin{equation}
P_{\rm st}(x) =  Z^{-1} \ \rme^{- \frac{1}{D} \ V(x)} = P_{\rm GB}(x)
\end{equation}
independently of $\alpha$ and $g$. $P_{\rm GB}$ stands for the Gibbs-Boltzmann equilibrium distribution function 
in the canonical ensemble of statistical mechanics.

Therefore, in order to describe the Markovian stochastic dynamics of a physical quantity subject to multiplicative white noise and that reaches a usual Gibbs-Boltzmann equilibrium measure, one needs to work with the drifted Langevin equation
\begin{equation}
\rmd_t x(t) = -g^2(x) V'(x) + 2D(1-\alpha) g(x) g'(x) + g(x) \xi(t)  
 \label{eq:x-eom-drifted}
\end{equation}
in a generic $\alpha$-prescription.
Several features of this equation should be remarked. First, there is a non-trivial additional `drift force'
even in the Stratonovich mid-point ($\alpha=1/2$) prescription.  Second, the additional term {\it is not} equal to 
the one in the chain rule~(\ref{eq:x-chain}). 
Moreover, this equation is equivalent to the original `undrifted' Eq.~(\ref{eq:x-eom}) with a post-point prescription $\alpha=1$~\cite{Hanggi78,Hanggi80,Hanggi1982,Klimontovich}.

Although it is conventional to work with Langevin equations of the form of Eq.~(\ref{eq:x-eom}) or Eq.~(\ref{eq:x-eom-drifted}) in which 
the left-hand side (lhs) is solely given by the time derivative, $\rmd_t x$, it can be illuminating to re-write the latter as
\begin{equation}
 k^2(x) \rmd_t x(t)  =  -  V'(x) -  2D  (1-\alpha) \partial_x \ln|k(x)|   + k(x) \xi(t)\;,
\label{eq:eq-a-la-Aron-Biroli-LFC}
\end{equation}
where we re-parametrised $g(x) \equiv 1/k(x)$. Equation~(\ref{eq:eq-a-la-Aron-Biroli-LFC}) has the form claimed in Ref.~\cite{AronLeticia2010} for the Markovian overdamped dynamics of particles subject to forces deriving from a potential $V$ and interacting with a bath of oscillators \textit{via} a non-linear coupling $K(x)$ with $K'(x)\equiv k(x)$. The exact  integration over the degrees of freedom of the bath gives rise to a viscous friction force, here in the left-hand side (lhs), as well as the  multiplicative noise in the right-hand side (rhs). Note that the drift force was not discussed in Ref.~\cite{AronLeticia2010} since the focus in this paper was on non-Markovian dynamics (either because of the presence of a coloured noise or inertia) for which case no drift force is needed to ensure the convergence to the usual Gibbs-Boltzmann distribution. 
We can therefore re-interpret the time derivative, $\rmd_t x$, in the lhs of Eq.~(\ref{eq:x-eom}) or Eq.~(\ref{eq:x-eom-drifted}) as originating from the dissipative interaction with the same bath that is responsible for the random noise $\xi(t)$.

\subsection{The path-integral formulation}
\label{subsec:path-int}
{
The stochastic dynamics of Markov processes governed by Langevin equations can be formulated in terms of path integrals. This approach has been first developed for cases with an additive noise~\cite{Onsager53,Graham73,Martin73,Janssen76,deDominicis76,Pythian77}.
It was later generalised to cases with a multiplicative noise~\cite{Leschke77} and extended to various discretization schemes and to higher dimensions~\cite{Langouche79,Langouche81}. 
 Below, we recall the construction of such a path-integral representation on the case of the multiplicative-noise equation~(\ref{eq:x-eom}) by following a procedure \textit{\`a la} Martin-Siggia-Rose-Janssen-deDominicis (MSRJD).}

The probability distribution for a given trajectory of $x$, with initial condition $x(-{\cal T})$ at time $-{\cal T}$ 
distributed according to $P_{\rm i}(x(-{\cal T}))$, and governed by Eq.~(\ref{eq:x-eom}) is 
\begin{eqnarray}
 P[x; \alpha]
  &\propto& P_{\rm i}(x_{-{\cal T}}) \ \langle \, |\mathcal{J}| \ \int\uD[\hat{x}] \ \rme^{\int  \rmi\hat{x}_t\,  {\rm Eq}_t[x,\xi; \alpha]} \, \rangle
 \label{eq:P(x)}
\end{eqnarray}
where we used the short-hand notation $\int$ for the time integral in the exponential that runs over the 
(symmetrised for convenience) time interval $[-{\cal T}, {\cal T}]$, and $x_t = x(t)$ for the 
time-dependent functions. The brackets denote the statistical average over all possible histories of the noise $\xi$.
We introduced the auxiliary field $\rmi\hat x_t$ to exponentiate the $\delta$-function that imposes that $x_t$ be the solution to the Langevin equation at all times (for a given history of $\xi$). We wrote the latter constraint in the compact form $\mbox{Eq}_t[x,\xi; \alpha]=0$. The
Jacobian ${\cal J}$ is given by
\begin{align}
 \mathcal{J} \equiv \det_{tt'} \left[ \frac{\delta \mbox{Eq}_t[x,\xi; \alpha]}{\delta x_{t'}} \right]
\; . 
\label{eq:J(x)}
\end{align}
The calculations detailed in App.~\ref{app} are similar to the ones in Refs.~\cite{Janssen79,Langouche79,Langouche81,Janssen92}, and yield for the case of the `undrifted' Langevin equation (\ref{eq:x-eom})
\begin{equation}
P[x; \alpha] \propto \int {\cal D}[\hat x] \ P[x,\rmi \hat x; \alpha]
\qquad
\mbox{with}
\qquad
P[x,\rmi \hat x;\alpha] =  \rme^{S[x,\rmi\hat x;\alpha]}
\label{eq:path-integral}
\end{equation}
and the Martin-Siggia-Rose-Janssen-deDominicis (MSRJD) action~\cite{Tirapegui-book,Lubensky2007}
\begin{eqnarray}
 S[x,\rmi\hat{x};\alpha]  
 \equiv\!\!
 \int \!\! \left[-\rmi\hat{x}_t ( \dot x_t \!-\! f_t \!+\! 2D\alpha g_t' g_t )
 \!+\! D ( \rmi\hat{x}_t )^2 g_t^2 \!-\! \alpha f_t' \right]
 \!+\! \ln P_{\rm i}(x_{-{\cal T}}).
 \label{eq:x-action-1}
\end{eqnarray}
This action coincides with the form given in Refs.~\cite{Leschke77, Lubensky2007}. It differs from the one in Ref.~\cite{Jarzynski08} since the authors used a post-point prescription in the stochastic equation (\textit{i.e.} $\alpha = 1$) while using a mid-point Stratonovitch prescription in the construction of the path integral formalism.
We disagree with the statements made in Refs.~\cite{Ao14a,Ao14b} concerning the invalidity of action functional in Ref.~\cite{Lubensky2007}. 
Note that the use of a non-linear change of variable within the path integral is known to be problematic, even when starting from a mid-point discretisation, unless the underlying discretisation is treated with great care. The non-trivial effects of non-linear transformations were already observed in a quantum field theory context~\cite{Gervais76,Salomonson77,Alfaro90,Apfeldorf96} and they appear within stochastic field theory as well~\cite{Langouche79,Tirapegui-book}. In other words, covariance of the action functional under general coordinate transformations comes with highly non-trivial treatment of the underlying discretization prescription.

With the addition of the drift force that ensures the approach to the usual Gibbs-Boltz\-mann equilibrium, see
Eq.~(\ref{eq:equil-choice-f}), the MSRJD action reads
\begin{align}
 S[x,\rmi\hat{x};\alpha]  & = 
 \int \left\{ -\rmi\hat{x}_t [ \dot x_t + g_t^2 V_t' - 2D (1-2\alpha) g_t' g_t ]
 + D ( \rmi\hat{x}_t )^2 g_t^2 
 \right.
 \nonumber\\
 & \qquad\;\;\;\;\;
 \left.
 - \alpha \partial_x [ -g_t^2 V_t' + 2 D (1-\alpha) g_t g_t'] \right\} 
 + \ln P_{\rmi}(x_{-{\cal T}})
 \label{eq:x-action-1-repeat}
 \; . 
\end{align}

\subsection{Fluctuations}
\label{sec:transformations}

The stochastic nature of the dynamics is responsible for fluctuations of the field (here $x_t$) and more generally of all the possible physical observables that 
depend on this field [\textit{i.e.}, any $A(x_t)$]. Amongst the few universal results that apply to these dynamics, there is a class of 
exact relations between the path probabilities that are very precious since they lead to strong relations between observables.
In a functional formalism, these relations can be proven by 1) making use of physical symmetries or broken symmetries of the system or its dynamics, 2) exploiting the invariance of the generating functional under a dummy linear change of integration variables.
In this Subsection, we shall discuss these relations in the context of stochastic Markov processes with multiplicative white noise such as the ones defined by Eq.~(\ref{eq:x-eom-drifted}).

\subsubsection{Relation between path probabilities}

Let us consider the cases in which the force $f$ depends on a set of externally controlled, possibly time-dependent, parameters $\lambda_t$.
The stochastic process is characterised by the path integral (\ref{eq:path-integral}) that expresses the joint probability distribution, 
$P[x, \rmi \hat x;\alpha, \lambda]$,  of the 
time series $\{ x_t, \ \rmi \hat x_t\}$ of the physical and the auxiliary 
fields, in the $\alpha$-prescription, and under the set of parameters $\lambda_t$. 
Following Crooks~\cite{Crooks00}, we ask how does $P[x, \rmi \hat x; \alpha, \lambda]$ compare to the 
probability distribution of the transformed time-dependent
variables $\{{\cal T} x_t,  {\cal T} \rmi \hat x_t \}$ in another discretisation prescription, $\overline \alpha$, 
and, possibly, under a transformed set of parameters, $\overline \lambda_t$. 
By choosing adequately the 
transformation rules ${\cal T} x$, ${\cal T} \rmi \hat x$, $\overline \alpha$ 
and $\overline \lambda$
we will obtain relations of the type
\begin{eqnarray}
\frac{P[{\cal T} x,  {\cal T} \rmi \hat x; \overline \alpha, \overline \lambda]}{P[x, \rmi \hat x; \alpha, \lambda]} = 
\rme^{ \Delta S[x, \rmi \hat x; \alpha, \lambda]}
\;.
\label{eq:identity}
\end{eqnarray}
We have distinguished the notation for the transformation of the dynamical fields, ${\cal T} x,  {\cal T} \rmi \hat x$, from the changes in the discretisation parameter, $\overline \alpha$, and 
the external time-dependent parameter, $\overline \lambda_t$.

The relation (\ref{eq:identity})  implies, for the average of a generic function $A$ of the physical and auxiliary fields (but, for simplicity, not of their time derivatives)
\begin{eqnarray}
&&
\int {\cal D}[x,  \hat x] \ A[ x,  \rmi \hat x] \ 
P[ x, \rmi \hat x; \overline \alpha,\overline\lambda]
\nonumber\\
&& 
\qquad
=
\int {\cal D}[{\cal T} x,{\cal T}  \hat x] \ A[{\cal T} x, {\cal T} \rmi \hat x] \ 
P[{\cal T} x, {\cal T} \rmi \hat x; \overline \alpha,\overline\lambda]
\nonumber\\
&& 
\qquad
=
\int {\cal D}[{\cal T} x,{\cal T} \hat x] \ A[{\cal T} x, {\cal T} \rmi \hat x] \ 
P[x, \rmi \hat x; \alpha, \lambda] \ \rme^{\Delta S[x, \rmi \hat x; \alpha, \lambda]}
\; . 
\end{eqnarray}
Moreover, if the measure over the transformed fields can be related to the one over the original ones with a unit Jacobian, and if the domain of integration at each time slice, here the real axis, is unchanged or can be taken back to the real axis, then the relation above becomes
\begin{eqnarray}
&&
\int {\cal D}[x,  \hat x] \ A[ x,  \rmi \hat x] \ 
P[ x, \rmi \hat x; \overline \alpha,\overline\lambda]
\nonumber\\
&& 
\qquad\quad
=
\int {\cal D}[ x, \hat x] \ A[{\cal T} x, {\cal T} \rmi \hat x] 
\  \rme^{\Delta S[x, \rmi \hat x; \alpha, \lambda]} \ P[x, \rmi \hat x, \alpha, \lambda] 
\; . 
\label{eq:relation-paths}
\end{eqnarray}
Writing $A[{\cal T} x, {\cal T} \rmi \hat x]$ as a new function of the original 
fields $x$ and $\rmi \hat x$, say $B[x, \rmi \hat x] \equiv
A[{\cal T} x, {\cal T} \rmi \hat x]$, one has a generic relation between averages of different functions:
\begin{eqnarray}
&&
\int {\cal D}[x,  \hat x] \ A[ x,  \rmi \hat x] \ 
P[ x, \rmi \hat x; \overline \alpha,\overline\lambda]
\nonumber\\
&& 
\qquad\quad
=
\int {\cal D}[ x, \hat x] \ B[x, \rmi \hat x] \ \rme^{\Delta S[x, \rmi \hat x; \alpha, \lambda]}
\ P[x, \rmi \hat x; \alpha, \lambda] 
\; . 
\label{eq:relation-paths-bis}
\end{eqnarray}
With different choices of the function $A$, and their associated $B$, one can derive various relation.
In particular, 
choosing $A=1$, 
\begin{eqnarray}
&&
1 = \langle \rme^{\Delta S[x, \rmi \hat x; \alpha, \lambda]} \rangle
\; . 
\label{eq:uno}
\end{eqnarray}

In Sec.~\ref{subsec:transformation}, we will identify the transformation ${\cal T}_{\rm eq}$, associated with the time-reversal invariance of equilibrium dynamics, that leaves the 
probability density invariant ($\Delta S = 0$) whenever the system is subject to equilibrium conditions, meaning initial conditions drawn from the Gibbs-Boltzmann distribution 
$P_{\rm GB}\propto \rme^{-V/D}$ and dynamics given by the Langevin equation Eq.~(\ref{eq:x-eom-drifted}), with a drift force deriving from the same confining  potential $V$, and in contact with a thermal bath at the same temperature such that $k_BT = D$.
We later use this invariance to derive generic properties of equilibrium dynamics, 
such as the fluctuation-dissipation theorem (Sec.~\ref{subsec:FDT}). Out of equilibrium, $\Delta S \neq 0$, and we will derive in Sec.~\ref{subsec:FT} 
various fluctuation relations that have been extensively studied in 
recent years~\cite{Jarzynski97,Kurchan98,Lebowitz99,Evans02,Ritort03, Maes03,Park04,Bustamante05,seifert2008,Zamponi}. 

\subsubsection{The time-reversal transformation}
\label{subsec:transformation}

We look for the invariance of the generating functional that 
corresponds to the  time-reversal invariance of the equilibrium dynamics. 
For a clear discussion of time-reversal in the context of Markovian equations of motion, see Refs.~\cite{Haussmann86,Nelson67}.
Although the action functional in Eq.~(\ref{eq:x-action-1-repeat}) is relatively cumbersome, 
the identification of the correct field transformation that leaves it invariant can be simplified by the 
fact that one expects the time-reversal invariance to hold for the system and its environment separately.
In other words, we expect the terms in the action that have their origin in the coupling to the bath to transform 
independently from the rest of the action. {We identify them, see the discussion around Eq.~(\ref{eq:eq-a-la-Aron-Biroli-LFC}), and collect them in}
\begin{equation}
\\
S_{\rm diss} [x,\rmi\hat{x}] \equiv \int  \rmi\hat x_t  [ D  \rmi\hat{x}_t  g_t^2 - {\rm d}_t^{(\alpha)} x_t   ]
\label{eq:S-diss}
\end{equation}
where, to simplify notations, we defined
\begin{eqnarray}
{\rm d}_t^{(\alpha)}x_t &\equiv& {\rm d}_t x_t - 2 D (1-2\alpha) g_tg_t'
\label{eq:covariant-derivative}
\end{eqnarray}
and, we recall, $D= k_BT = \beta^{-1}$.
For a field $x_t$ corresponding to a physical quantity $x$ that is even under time-reversal transformation (such as the particle's position), 
the transformation of the physical field must naturally be $x_t \mapsto x_{-t}$.
The expression of $S_{\rm diss}$ in Eq.(\ref{eq:S-diss}) suggests that we look for a transformation such that ${\rm d}_t^{(\alpha)} x_t$ behaves as a usual time derivative 
under time reversal, \textit{i.e.}
\begin{equation}
{\rm d}_t^{(\alpha)} x_t  \mapsto  -{\rm d}_{-t}^{(\alpha)} x_{-t}
\;.
\end{equation}
This is only true if we  simultaneously transform  the discretisation parameter $\alpha \mapsto 1 - \alpha$.
Altogether, we are led to propose the following transformation of the dynamical field $x_t$ and its associated auxiliary field $\rmi \hat x_t$
\begin{eqnarray} \label{eq:Teq}
{\cal T}_{\rm eq} = 
\left\{
\begin{array}{rcl}
x_t &\mapsto& x_{-t}
\; , 
\label{eq:time-rev-x-alpha}
\\
\rmi \hat x_t &\mapsto & \rmi \hat x_{-t} - D^{-1} g^{-2}_{-t} {\rm d}_{-t}^{(\alpha)} x_{-t} 
\;,
\end{array}
\right.
\end{eqnarray}
complemented with the transformation of the  discretisation parameter
\begin{align}
\alpha & \mapsto \overline\alpha \equiv 1- \alpha\;.
\end{align}

It is easy to check that $S_{\rm diss}$ is indeed invariant under this transformation:
\begin{align}
S_{\rm diss}[{\cal T}_{\rm eq}  x, {\cal T}_{\rm eq} \rmi \hat x; \overline \alpha] 
&= 
\int 
\left\{ 
[\rmi \hat x_{-t} - D^{-1} g^{-2}_{-t} {\rm d}^{(\alpha)}_{-t} x_{-t} ]
\right.
\nonumber\\
&
\qquad\quad
\left.
\times
[   {\rm d}^{(\alpha)}_{-t} x _{-t} + D \  g_{-t}^2 \rmi \hat x_{-t} - D g_{-t}^2 \beta g^{-2}_{-t} {\rm d}^{(\alpha)}_{-t} x_{-t} ]
\right\}
\nonumber\\
&
= 
\int 
\left\{ 
[\rmi \hat x_{-t} - D^{-1} g^{-2}_{-t} {\rm d}^{(\alpha)}_{-t} x_{-t} ] \
D \  g_{-t}^2 \rmi \hat x_{-t} 
\right\}
\nonumber\\
&
= 
\int 
\left\{ 
[ D \  g_{t}^2 \rmi \hat x_{t} -  {\rm d}^{(\alpha)}_t x_{t} ] \
\rmi \hat x_{t} 
\right\} 
 = 
S_{\rm diss}[x,\rmi \hat x; \alpha]
\; .
\end{align}

We now have to check that the remaining terms in the action functional are also invariant under the proposed transformation. In the potential case with no time-dependent 
parameter ($\partial_t \lambda_t= 0$)
and a drift force ensuring the convergence to the usual Gibbs-Boltzmann equilibrium measure, the remaining terms are gathered into
\begin{eqnarray*}
S_{{\rm det}+{\rm jac}}[x,\rmi\hat{x}; \alpha]  = 
 \ln P_{\rmi}(x_{-{\cal T}}) - \! \int   \rmi\hat{x}  g_t^2 V_t' 
 - \alpha  \! \int \partial_x [ -g_t^2 V_t' + 2 D (1-\alpha) g_t g_t'] 
 \qquad
\label{eq:S-jac}
\end{eqnarray*}
and they transform as
\begin{eqnarray}
&&
S_{{\rm det}+{\rm jac}}[ {\cal T}_{\rm eq}  x,  {\cal T}_{\rm eq} \rmi \hat x; \overline \alpha] 
\nonumber\\
&& 
\quad
= 
 \ln P_{\rm i}(x_{\cal T}) +
\int \left[ 
(-\rmi \hat x_{-t} + D^{-1} g^{-2}_{-t} {\rm d}_{-t}^{(\alpha)} x_{-t} )
g^2_{-t} V'_{-t}
+ (1-\alpha)  
 \partial_{x_{-t}} (g^2_{-t} V'_{-t}) 
 \right]
 \nonumber\\
&& \qquad\qquad\qquad\qquad
-2D \alpha (1-\alpha)  \int \partial_x (g_{-t} g_{-t}')
 \nonumber\\
&&
\quad 
=  \ln P_{\rm i}(x_{\cal T})
+
\int \left[ 
-\rmi \hat x_{t} g^2_t V'_{t} + D^{-1} {\rm d}_t^{(\alpha)} x_{t} V'_{t}
+ (1-\alpha) 
 \partial_{x_{t}} (g^2_{t} V'_{t}) 
 \right]
 \nonumber\\
&& \qquad\qquad\qquad\qquad
-2D \alpha (1-\alpha) \int \partial_x (g_{t} g_{t}')
 \; . 
\end{eqnarray}
We recognise that the first boundary term, $\ln P_{\rm i}(x_{\cal T})$, needs to be taken back 
to the initial time, $-{\cal T}$, if one wants to recover the original $S_{{\rm det}+{\rm jac}}[\rmi \hat x, x;\alpha]$; the second and last terms are already part of $S_{{\rm det}+{\rm jac}}[\rmi \hat x, x;\alpha]$; rewriting  $1-\alpha = \alpha + (1-2\alpha)$, the fourth term produces the last piece needed to fully reconstruct the original $S_{{\rm det}+{\rm jac}}[\rmi \hat x, x;\alpha]$.
All in all, we have
\begin{align}
 & S_{{\rm det}+{\rm jac}}[ {\cal T}_{\rm eq} x, {\cal T}_{\rm eq} \rmi \hat  x; \overline \alpha] 
 =  \ S_{{\rm det}+{\rm jac}}[ x,  \rmi \hat x, \alpha] 
+ D^{-1} \int {\rm d}_t^{(\alpha)} x_{t}  \ V'_{t}
\\
& \qquad\qquad\qquad 
+ (1-2\alpha) \int \partial_{x_{t}} (g^2_{t} V'_{t}) 
 + \ln P_{\rm i}(x_{\cal T}) - \ln P_{\rm i}(x_{-{\cal T}})
 \; .  \nonumber
\end{align}
Using the explicit form of ${\rm d}_t^{(\alpha)} x_t$ in Eq.~(\ref{eq:covariant-derivative}),
we can simplify this expression as follows
\begin{align}
S_{{\rm det}+{\rm jac}}[ {\cal T}_{\rm eq}  x, {\cal T}_{\rm eq} \rmi \hat x; \overline \alpha] 
= & \ S_{{\rm det}+{\rm jac}}[x,  \rmi \hat  x; \alpha] 
+ D^{-1} \int \rmd_t x_{t}  \ V'_{t}  \label{eq:action-in-transformed}
\\
& + (1-2\alpha) \int g^2_{t} \ V''_{t} 
 + \ln P_{\rm i}(x_{\cal T}) - \ln P_{\rm i}(x_{-{\cal T}})
 \; .  \nonumber
\end{align}
Replacing the term in $\rmd_t x_t \ V_t'$ above by using the chain rule of stochastic calculus recalled in Eq.~(\ref{eq:x-chain})~\cite{arenas2012}, 
\begin{equation}
\rmd_t V_t = \rmd_t x_t \ V_t' + (1-2\alpha) D g_t^2 \ V_t'',
\label{eq:time derivative-V}
\end{equation}
we obtain
\begin{align}
 S_{{\rm det}+{\rm jac}}[ {\cal T}_{\rm eq}  x, {\cal T}_{\rm eq} \rmi \hat x; \overline\alpha] 
=& \ S_{{\rm det}+{\rm jac}}[  x, \rmi \hat x; \alpha] 
\nonumber\\
& 
 + D^{-1} \int \rmd_t V_t
+ \ln P_{\rm i}(x_{\cal T}) - \ln P_{\rm i}(x_{-{\cal T}})
 \; . 
\end{align}
Finally, with initial conditions drawn from the Gibbs-Boltzmann distribution
\begin{equation}
P_{\rm i}(x_{-{\cal T}}) = Z^{-1}  \  \rme^{-D^{-1} V(x_{-{\cal T}})}\,,
\end{equation}
we end the proof of the full invariance of the equilibrium action functional in Eq.~(\ref{eq:x-action-1-repeat}) under the transformation ${\cal T}_{\rm eq}$ given in Eq.~(\ref{eq:Teq}):
 \begin{eqnarray}
S[ {\cal T}_{\rm eq}  x, {\cal T}_{\rm eq} \rmi \hat x; \overline\alpha] 
 = S[  x, \rmi \hat x; \alpha] 
\end{eqnarray}
and $\Delta S=0$.
Note that to achieve this invariance, there was a subtle interplay between the contributions coming from the deterministic part of the action and the ones 
coming from the $\alpha$-dependent Jacobian.

This invariance of the action functional yields the following relation between path probabilities
\begin{equation}
P[x, \rmi \hat x ;  \alpha] \  \uD{[x,\hat x]}
= P[{\cal T}_{\rm eq} x, {\cal T}_{\rm eq} \rmi \hat x ; \overline \alpha] \ \uD{[ {\cal T}_{\rm eq} x,{\cal T}_{\rm eq} \rmi \hat x]}
\; . 
\end{equation}
After the transformation ${\cal T}_{\rm eq}$, the domain of integration of $\hat x_t $ at each time slice of the generating functional is shifted from the real axis to the complex 
line with a constant imaginary part $\rmi D^{-1} \rmd_t x_t$. Using the analyticity of $\exp S[x;\rmi\hat x; \alpha] $, one can return to an integration over the real axis by closing the 
contour at both infinities and by dropping the contributions of the vertical ends that vanish owing to the term $D (\rmi\hat x_t)^2$.
Note also that the Jacobian associated to the change of variables $\{ x, \rmi \hat x\} \mapsto 
\{ {\cal T}_{\rm eq} x, {\cal T}_{\rm eq} \rmi \hat x \}$ is unity. Finally, $\uD{[x,\hat x]} =  \uD{[ {\cal T}_{\rm eq} x,{\cal T}_{\rm eq} \rmi \hat x]}$ and we obtain the following relation between the forward and backward path probabilities
\begin{align}
\frac{P_{\rm B}[x, \rmi \hat x]}{P_{\rm F}[x, \rmi \hat x]} = 1\;, \label{eq:equal-prob}
\end{align}
where we defined 
\begin{align}
P_{\rm F}[x, \rmi \hat x] \equiv  P[x, \rmi \hat x, \alpha]\;, \qquad\qquad
P_{\rm B}[x, \rmi \hat x] \equiv  P[{\cal T}_{\rm eq} x, {\cal T}_{\rm eq} \rmi \hat x,\bar \alpha]\;.
\end{align}
The relation (\ref{eq:equal-prob})  is valid whenever the system is in thermal equilibrium.

\subsubsection{The fluctuation-dissipation theorem}
\label{subsec:FDT}

The fluctuation-dissipation theorem~\cite{Callen51,Kubo57,Kubo66,Groot,StratoBookI,KuboBookII} is a model-independent relation between the linear response and the correlation of spontaneous equilibrium fluctuations of a given observable.
The linear response of $x$ with respect to a previous perturbation  is defined as 
\begin{equation}
R_\alpha(t,t') = \left. \frac{\delta \langle x(t)\rangle_h}{\delta h(t')} \right|_{h=0}
\; , 
\end{equation}
where the infinitesimal perturbation $h$ couples linearly to the field $x$ in such a way that the potential $V\to V_h = V - h x$
and, therefore, $V'\to V'_h = V' - h$. 
In the path-integral formulation, the linear response is given by  
\begin{equation}
R_\alpha(t,t') = \int {\cal D}[x, \rmi \hat x] \ x_t \left. \frac{\delta S_h[x,\rmi \hat x; \alpha]}{\delta h_{t'}} \right|_{h=0} \ \rme^{S[x,\rmi \hat x; \alpha]}
\end{equation}
where the action has been modified as
\begin{equation}
S_h[x, \rmi \hat x; \alpha] = S[x,\rmi \hat x; \alpha] + \int  h_t \left[ \rmi\hat x_t g_t^2 - 2 \alpha g_t g'_t \right]
\;  .
\end{equation}
Therefore, the linear response is expressed as a correlation function reading 
\begin{equation}
R_\alpha(t,t') =  \langle x_t [\rmi \hat x_{t'} g_{t'}^2 - 2\alpha g_{t'} g'_{t'} ]\rangle_{S[x,\rmi \hat x; \alpha]}
\label{eq:linear-response}
\end{equation}
where the average has to be taken with the  measure given by the unperturbed action $S[x,\rmi \hat x; \alpha]$.
The subindex $\alpha$ expresses the fact that the stochastic process is defined with a discretisation parameter  $\alpha$. Exchanging momentarily $\alpha$ by $1-\alpha$ one has
\begin{equation}
R_{1-\alpha}(t,t') =  \langle x_t [\rmi \hat x_{t'} g_{t'}^2 - 2(1-\alpha) g_{t'} g'_{t'} ]\rangle_{S[x,\rmi \hat x; 1-\alpha]}\;.
\label{eq:linear-response-1-alpha}
\end{equation}
Take now the expression in Eq.~(\ref{eq:linear-response-1-alpha}) and perform the change variables in the path integral from $\{ x_t, \rmi \hat x_t \}$ to $\{ {\cal T}_{\rm eq} x_t, {\cal T}_{\rm eq} \rmi \hat x_t \}$:
\begin{equation}
R_{1-\alpha}(t,t') = 
 \langle {\cal T}_ c x_t [{\cal T}_{\rm eq} \rmi \hat x_{t'} {\cal T}_{\rm eq} g_{t'}^2 - 2(1-\alpha) {\cal T}_{\rm eq} g_{t'} {\cal T}_{\rm eq} g'_{t'} ]\rangle_{S[{\cal T}_{\rm eq} x, {\cal T}_{\rm eq} \rmi \hat x; 1-\alpha]}
\label{eq:linear-response1}
\end{equation}
where ${\cal T}_{\rm eq} g = g({\cal T}_{\rm eq} x_t)$ and similarly for ${\cal T}_{\rm eq} g'$.
Using that $S[{\cal T}_{\rm eq} x, {\cal T}_{\rm eq} \rmi \hat x; 1-\alpha] = S[x, \rmi \hat x; \alpha]$,
and applying the transformation ${\cal T}_{\rm eq}$ defined in Eq.~(\ref{eq:Teq}) to the function of $x_t$ and $\rmi \hat x_{t'}$ to 
be averaged, one has 
\begin{eqnarray}
&&
R_{1-\alpha}(t,t') = 
 \langle x_{-t} \left\{ [\rmi \hat x_{-t'}  - D^{-1}  g^{-2}_{-t'} {\rm d}_{-t'}^{(\alpha)} x_{-t'}] g^2_{-t'} 
\right.
\nonumber\\
&& 
\qquad \qquad \qquad \;\;\;
\left.
- 2(1-\alpha) g_{-t'} g'_{-t'}] 
\right\}
\rangle_{S[\rmi \hat x, x; \alpha]}
\nonumber\\
&& 
\qquad \qquad\;\;\,
= 
 \langle x_{-t} [\rmi \hat x_{-t'}  g^2_{-t'} - 2(1-\alpha) g_{-t'} g'_{-t'} \rangle_{S[\rmi \hat x, x; \alpha]}
\nonumber\\
&& 
\qquad\qquad\;\;\;\;\;\;\;
- D^{-1}  \langle x_{-t} {\rm d}_{-t'}^{(\alpha)} x_{-t'}]  \rangle_{S[\rmi \hat x, x; \alpha]}
\nonumber\\
&& 
\qquad \qquad\;\;\,
= 
 \langle x_{-t} [\rmi \hat x_{-t'}  g^2_{-t'} - 2\alpha g_{-t'} g'_{-t'} \rangle_{S[\rmi \hat x, x; \alpha]}
\nonumber\\
&& 
\qquad\qquad\;\;\;\;\;\;\;
- \langle x_{-t}  2(1-2\alpha) g_{-t'} g'_{-t'} \rangle_{S[\rmi \hat x, x; \alpha]}
- D^{-1}  \langle x_{-t} {\rm d}_{-t'}^{(\alpha)} x_{-t'}]  \rangle_{S[\rmi \hat x, x; \alpha]}
\nonumber
\end{eqnarray}
Identifying $R_\alpha(-t,-t')$ in the first term in the rhs
and using now ${\rm d}_{-t}^{(\alpha)} x_{-t} = \rmd_{-t} x_{-t} - 2D (1-2\alpha) g_{-t} g'_{-t}$,
\begin{eqnarray}
&&
R_{1-\alpha}(t,t') 
= 
R_\alpha(-t,-t') + D^{-1} \langle x_{-t} \rmd_{t'} x_{-t'}\rangle_{S[\rmi \hat x, x; \alpha]}
\; . 
\end{eqnarray}
Using the fact that the physics cannot depend on the discretisation parameter [see the discussion below 
the drifted Fokker-Planck Eq.~(\ref{eq:FP-correct})], we can drop the irrelevant index $\alpha$ (or $1-\alpha$) in the linear response and the correlation function and 
\begin{eqnarray}
&&
R(t,t') 
= 
R(-t,-t')
+ D^{-1} \partial_{t'} C(-t,-t')
\; .
\end{eqnarray}
We apply the transformation ${\cal T}_{\rm eq}$  once again on the correlation function in the rhs to show
$C(-t,-t') = C(t,t')$. Owing to the time-translational invariance of equilibrium dynamics, $C(t,t') = C(\tau)$ and $R(t,t') = R(\tau)$ where $\tau \equiv t-t'$, and to the causality of the response $R(\tau) =0 $ for $\tau <0$, we obtain the celebrated fluctuation-dissipation theorem (FDT)
\begin{align}
R(\tau) = - D^{-1} \Theta(\tau) \rmd_\tau C(\tau) = - \beta \Theta(\tau) \rmd_\tau C(\tau)
\; . 
\end{align}
Here $\Theta(\tau)$ is the Heaviside step function.

\subsubsection{Broken symmetry and fluctuation theorems}
\label{subsec:FT}

There are various ways to drive a system out of equilibrium, \textit{e.g.}, by changing an external parameter in the potential in time, 
$\lambda_t$,  or by using initial conditions that are not in equilibrium, $P_{\rm i}(x_{-{\cal T}}) \neq P_{\rm GB}(x_{-{\cal T}})$.
In these cases, the associated time-reversal symmetry of the dynamics is broken and 
the action functional is no longer invariant under the field transformation $\mathcal{T}_{\rm eq}$. In practice, this means that $\Delta S$ defined in Eq.~(\ref{eq:identity}) does not vanish and the very same transformation technique that we used earlier can now be used to derive exact out-of-equilibrium relations between path probabilities, the so-called fluctuation relations.

The time-reversed dynamics corresponds to evaluating the action functional in the transformed fields ${\cal T}_{\rm eq} x_t $ and ${\cal T}_{\rm eq} \rmi \hat x_t$, the discretisation prescription parameter $\overline{\alpha} = 1 - \alpha$, and the time-reversed protocol
\begin{equation}
\overline \lambda_t = \lambda_{-t}\;.
\end{equation}
In order to evaluate $\Delta S$, one first notices that the dissipative part of the action $S_{\rm diss}$ does not depend upon the applied force 
nor the initial condition. Therefore, it remains invariant under ${\cal T}_{\rm eq}$. However, as $S_{{\rm det} + {\rm jac}}$ depends on both the initial distribution and the force, we expect it to yield  $\Delta S \neq 0$. More precisely,  the terms contributing to $\Delta S$ are given in Eq.~(\ref{eq:action-in-transformed}) where $V'_t$  and
$V_t''$ are  the first and second derivative 
with respect to the variable $x$ of the potential $V$,  $V'_t = \partial_x V_t$ and $V''_t = \partial_x^2 V$, respectively. If the potential 
depends on the time-dependent parameter $\lambda_t$, its total time derivative 
expressed in Eq.~(\ref{eq:time derivative-V}) acquires an extra term,
\begin{equation}
\rmd_t V_t = \rmd_t \lambda \ \partial_\lambda V_t + \rmd_t x \ \partial_x V_t  + (1-2\alpha) Dg_t^2 \ \partial_x^2 V_t
\; , 
\label{eq:time derivative-V-lambda}
\end{equation}
and we use this new relation to replace $\rmd_t x \ \partial_x V_t$ in the first term in Eq.~(\ref{eq:action-in-transformed}):
\begin{eqnarray}
 && S_{{\rm det}+{\rm jac}}[ {\cal T}_{\rm eq} \rmi \hat x, {\cal T}_{\rm eq} x; \overline \alpha, \overline \lambda] 
= S_{{\rm det}+{\rm jac}}[ \rmi \hat x, x; \alpha, \lambda] 
\nonumber\\
&& 
\qquad\qquad
 + \beta \int \rmd_t V_t - \beta \int \rmd_t \lambda_t \ \partial_\lambda V_t
+ \ln P_{\rm i}(x_{\cal T}, \lambda_{{\cal T}}) - \ln P_{\rm i}(x_{-{\cal T}}, \lambda_{-{\cal T}})
 \nonumber \\
 && 
\quad
=
S_{{\rm det}+{\rm jac}}[ \rmi \hat x, x; \alpha, \lambda] - D^{-1} \int \rmd_t \lambda_t \ \partial_\lambda V_t
+ D^{-1} V(x_{\cal T}, \lambda_{\cal T})
\nonumber\\
&&
\qquad\qquad
 - D^{-1} V(x_{-{\cal T}}, \lambda_{-{\cal T}}) 
+ \ln P_{\rm i}(x_{\cal T}, \lambda_{{\cal T}}) - \ln P_{\rm i}(x_{-{\cal T}}, \lambda_{-{\cal T}})
\; . 
\label{eq:transfoSdetjac1}
 \end{eqnarray}
Here, we made explicit the dependence on the time-dependent parameter $\lambda_t$ 
 of the potential and the initial probability distribution function. 
  The second term in the rhs  is related to  the work done by the time-dependent potential force
 \begin{equation}
 W =  \int {\rm d}_t \lambda_t \ \partial_\lambda V_t
 \; . 
 \end{equation}

 For generic $P_{\rm i}$ we cannot simplify further the last 
 four terms in the rhs of Eq.~(\ref{eq:transfoSdetjac1}) and  $\Delta S$ is the 
 stochastic entropy, defined as the sum of the Shannon entropy 
 [$\ln P_{\rm i}(x_{\cal T}, \lambda_{{\cal T}}) - \ln P_{\rm i}(x_{-{\cal T}})$] and
 the heat transfer ($\beta\mathcal{Q} = \beta\Delta V - \beta W$ and $\beta = D^{-1}$).
 
 If, instead, we assume that the system is initially prepared in the Gibbs-Boltzmann  
 distribution at temperature $k_{\rm B} T = D =  \beta^{-1}$, under a 
 potential \linebreak 
 $V(x_{-{\cal T}}, \lambda_{-{\cal T}})$, 
 \begin{equation}
 P_{\rm i}(x_{-{\cal T}}, \lambda_{-{\cal T}}) = Z^{-1}(\lambda_{-{\tau}}) \ \exp[-\beta V(x_{-{\cal T}}, \lambda_{-{\cal T}})]
 \; , 
 \end{equation} 
  we find
 \begin{align}
 S_{{\rm det}+{\rm jac}}[ {\cal T}_{\rm eq} \rmi \hat x, {\cal T}_{\rm eq} x;  \overline\alpha, \overline \lambda] 
 =& \
S_{{\rm det}+{\rm jac}}[ \rmi \hat x, x; \alpha, \lambda] - \beta \int \rmd_t \lambda_t \ \partial_\lambda V_t
\nonumber\\
&
- \ln Z(\lambda_{{\cal T}})  + \ln Z( \lambda_{-{\cal T}})
\; . 
 \end{align}
 The last two terms can be regrouped into
 \begin{equation}
\beta \Delta F  = \beta [F(\lambda_{{\cal T}}) - F(\lambda_{-{\cal T}})]= -\ln Z(\lambda_{{\cal T}})  + \ln Z( \lambda_{-{\cal T}})
\; , 
 \end{equation}
 the free-energy difference between the equilibrium state at the final and initial value of the parameter $\lambda_t$.
 Therefore, 
 \begin{eqnarray}
 S_{{\rm det}+{\rm jac}}[ {\cal T}_{\rm eq} \rmi \hat x, {\cal T}_{\rm eq} x; \overline\alpha, \overline \lambda] 
 &=&
S_{{\rm det}+{\rm jac}}[ \rmi \hat x, x; \alpha, \lambda] - \beta W
+\beta \Delta F
\;.
\end{eqnarray}
and ultimately
\begin{align}
\Delta S  & =   S_{{\rm det}+{\rm jac}}[ {\cal T}_{\rm eq} \rmi \hat x, {\cal T}_{\rm eq} x; \overline\alpha, \overline \lambda]  
- S_{{\rm det}+{\rm jac}}[ \rmi \hat x, x; \alpha, \lambda] \nonumber  \\
 & =
- \beta W +\beta \Delta F \; . 
\label{eq:DeltaS}
 \end{align}

In conclusion, we obtain the following relation between the forward and backward path probabilities
\begin{align}
\frac{P_{\rm B}[x,\rmi\hat x]}{P_{\rm F}[x,\rmi\hat x]} = \rme^{\Delta S}\;, \label{eq:relation-paths2}
\end{align}
where we defined
\begin{equation}
P_{\rm F}[x,\rmi\hat x] \equiv  P[x,\rmi\hat x ; \alpha,\lambda] \;, \quad\quad
P_{\rm B}[x,\rmi\hat x] \equiv  P[{\cal T}_{\rm eq} x,{\cal T}_{\rm eq} \rmi\hat x ; \bar \alpha, \bar \lambda]\;.  \label{eq:forw-back} 
\end{equation}

Multiplying both sides of Eq.~(\ref{eq:relation-paths2}) by $A$, a generic observable which can depend on $x$ and $\rmi \hat x$, 
and summing over all paths, one obtains
\begin{align}
\langle A(x, \rmi\hat x) \rangle_{\rm B} = \langle A(x, \rmi\hat x) \rme^{\Delta S} \rangle_{\rm F}\;,
\end{align}
where the subscripts ${\rm F}$ and ${\rm B}$ stand for averaging with the forward and backward path probability distributions defined in 
Eq.~(\ref{eq:forw-back}).

In particular, setting $A = 1$, one recovers the Jarzynski relation~\cite{Jarzynski97,Jarzynski04}
 \begin{equation}
 \langle \rme^{-\beta W} \rangle = \rme^{-\beta \Delta F} 
 \; . 
 \end{equation}
 Other fluctuation relations can be found by choosing other observables $A$.

\subsection{Schwinger-Dyson equations}

We end the analysis of the single-variable problem by presenting an easy derivation of the 
Schwinger-Dyson equations  {which govern the coupled dynamics of correlations and linear responses.} The proof is
based on the use of another set of transformation rules that leave the action and measure
invariant and hold in general.

\subsubsection{Out-of-equilibrium symmetry}

Let us consider the most generic out-of-equilibrium situation, \textit{i.e.} work with the original Langevin 
Eq.~(\ref{eq:x-eom}) without making  any assumption on the force $f$, that can possibly be time-dependent.
We recall that the MSRJD action functional associated to the dynamics reads
\begin{displaymath}
S[x,\rmi \hat x;\alpha] =  \int [- \rmi\hat x_t (\dot x_t - f_t + 2 D \alpha g_t g'_t  - D g_t^2 \rmi\hat x_t )- \alpha  f'_t ]
+ \ln P_{\rmi}(x_{-\mathcal{T}})
\;.
\end{displaymath}
This action is invariant under the  transformation~\cite{AronLeticia2010}
\begin{align}
\mathcal{T}_{\rm eom} \equiv 
\left\{
\begin{array}{rl}
x_t &\mapsto  \ \ x_t \; , \\
\rmi \hat x_t &\mapsto \displaystyle - \rmi \hat x_t + D^{-1} g_t^{-2} \left( \dot x_t - f_t + 2 \alpha D g_t g'_t \right)
\; , 
\end{array}
\right.
\end{align}
with no need to change the parameter $\alpha$.
We do not reproduce here the proof of invariance as it is rather straightforward.

\subsubsection{Ward-Takahashi identities}

Let us use this invariance of the action functional in the expression of the linear response to a perturbation $h$ such that $f \to f + g^2 h$
\begin{align}
R(t,t') &= \langle x_t (\rmi \hat x_{t'} g^2_{t'} - 2\alpha g_{t'} g'_{t'} ) \rangle \nonumber \\
&= -\langle x_t  \rmi \hat x_{t'} g^2_{t'}   \rangle
+ D^{-1} \langle x_t (\dot x_{t'} - f_{t'})
 \rangle \nonumber \\
&= - R(t,t')- 2 \alpha \langle x_t g_{t'}g'_{t'} \rangle
+ D^{-1} \langle x_t (\dot x_{t'} - f_{t'})   \rangle 
\; . 
\end{align}
We obtain
\begin{align}
2 D R(t,t') =  \partial_{t'} C(t,t') -  \langle x_t f_{t'} \rangle  - 2 \alpha D \langle x_t g_{t'}g'_{t'}  \rangle
\; . 
\end{align}
If one uses the drift force given in Eq.~(\ref{eq:equil-choice-f}) to ensure convergence to equilibrium with a Gibbs-Boltzmann measure $P_{\rm GB} \propto \rme^{-V/D}$, this yields
\begin{align}
2 D R(t,t') =  \partial_{t'} C(t,t') +  \langle x_t g^2_{t'} V'_{t'} \rangle  
- 2D \langle x_t g_{t'} g'_{t'}  \rangle
\; . 
\end{align}

\subsubsection{Composition of $\mathcal{T}_{\rm eq}$ and $\mathcal{T}_{\rm eom}$}

For equilibrium conditions, when working with the drifted Langevin Eq.~(\ref{eq:x-eom-drifted}), both transformations $\mathcal{T}_{\rm eq}$ and $\mathcal{T}_{\rm eom}$ 
are symmetries of the generating functional and therefore, their composition
\begin{align}
\mathcal{T}_{\rm eq} \circ \mathcal{T}_{\rm eom} \equiv 
\left\{
\begin{array}{rl}
x_t &\mapsto  x_{-t} \\
\rmi \hat x_t &\mapsto \displaystyle - \rmi \hat x_{-t} + D^{-1} V'_{-t} 
\end{array}
\right.
\end{align}
complemented with the transformation of the  discretisation parameter
\begin{align}
\alpha & \mapsto \overline\alpha \equiv 1- \alpha\;.
\end{align}
is also a symmetry of the generating functional.
Starting from the expression of the linear response, we derive
\begin{align}
R(t,t') &=  \langle x_t \rmi \hat x_{t'} \rangle - 2\alpha \langle x_t g_{t'} g'_{t'}  \rangle\nonumber \\
&=  \langle x_{-t} \left( - \rmi \hat x_{-t'} + D^{-1} V'_{-t'} \right) \rangle - 2(1-\alpha) \langle x_{-t} g_{-t'} g'_{-t'}  \rangle \nonumber \\
&= -R(-t,-t') + D^{-1} \langle x_{-t} V'_{-t'} \rangle - 2 \langle x_{-t} g_{-t'} g'_{-t'}  \rangle
\; .
\end{align}
Using the causality of the linear response, applying the transformation one more time, using the time-translational invariance of equilibrium dynamics,  and
$D=k_BT = \beta^{-1}$ we finally obtain the relation
\begin{align}
R(t-t') = \Theta(t-t')\left[\beta \langle x_t V'_{t'}  \rangle-2 \langle x_t  g_{t'} g'_{t'}  \rangle\right]\;.
\end{align}
This is a generalisation of the relations found in Refs.~\cite{Chatelain03, Ricci-Tersenghi03, Lippiello05}
for additive-noise stochastic processes, see also Ref.~\cite{Chetrite2011}. The special interest of this kind of relation is that it allows to compute the linear response, notably in a numerical evaluation, without any need to apply a perturbation,
by taking advantage of the expression of $R$ as the sum of two correlation terms.

 \section{Multi-variable stochastic Markov processes}
 \label{sec:LLG}
 
 In  Sec.~\ref{sec:one-dim}, we focused on a stochastic differential equation of a single variable 
 $x$. In a more general situation, the stochastic variable can be 
 a multi-dimensional vector. In this Section, we focus on the stochastic Landau-Lifshitz-Gilbert (sLLG) equation~\cite{Langevin-Coffey,Bertotti-etal}, a Langevin 
 equation describing the dynamics of a classical magnetic moment, a $3d$ vector $\mathbf{M}$, in contact with an 
 environment. 
 We analyse the time-reversal transformation of the magnetisation and the auxiliary vector  $\hat{\mathbf{M}}$  that leaves the action invariant under equilibrium conditions. 
We derive some of its consequences, such as the fluctuation-dissipation theorem. Furthermore, we also analyse the out-of-equilibrium dynamics 
driven by a spin-polarised current of electrons and we derive the corresponding fluctuation relations.
 
\subsection{The sLLG magnetisation dynamics}

In Ref.~\cite{Aron14}, we gave a detailed presentation of the stochastic Landau-Lifshitz-Gilbert (sLLG) equation 
that describes the dynamics of a magnetic moment $\mathbf{M}$ under the influence of a deterministic local magnetic field $\mathbf{H}_{\rm eff}$ and a thermal 
environment responsible for both dissipation and a fluctuating local magnetic field $\mathbf{H}(t)$.
Studying this equation in various discretisation prescriptions, we showed that unless the Stratonovich mid-point is used, a drift term is needed to 
conserve the magnetisation modulus in the course of time, $|\mathbf{M}| = M_s$, and to ensure the approach to Gibbs-Boltzmann equilibrium 
in the absence of non-conservative and time-dependent forces. Numerical checks of this fact were discussed in Ref.~\cite{Roma14}.
In Ref.~\cite{Aron14}, we also constructed the path-integral formalism
for the generating functional in the Cartesian and spherical coordinate systems. 
Adapting the results of Sec.~2, \textit{i.e.} identifying the field transformation that generalises the one in 
Eq.~(\ref{eq:time-rev-x-alpha}) to the physical problem at hand, we derive the corresponding fluctuation-dissipation theorem as well as the fluctuation relations. 


In the so-called Gilbert formulation, the sLLG equation reads~\cite{Stiles2006}
\begin{equation}
{\rm D}^{(\alpha)}_t \mathbf{M} =  -  \gamma_0 \mathbf{M} \wedge \left[ \mathbf{H}_{\eff} 
 + \mathbf{H}(t)
  - \frac{\eta}{M_s}  {\rm D}^{(\alpha)}_t \mathbf{M}
    \right]
    \; . 
 \label{eq:sLLG1-spintorque-adim}
\end{equation}
This equation has to be understood in the generic $\alpha$-prescription: 
$\overline{\mathbf{M}}_n \equiv \mathbf{M}_n + \alpha (\mathbf{M}_{n+1} -\mathbf{M}_n)$.
The $\alpha$-derivative is defined as
\begin{equation} \label{eq:derivcovariante}
{\rm D}_t^{(\alpha)} \equiv \rmd_t +  2D(1-2\alpha) \frac{ \gamma^2_0}{1+\eta^2\gamma_0^2}
\; , 
\end{equation}
and satisfies ${\rm D}_{-t}^{(1-\alpha)} = - {\rm D}_t^{(\alpha)}$. The second term in ${\mbox D}_t^{(\alpha)}$ 
is necessary to ensure
the conservation of the modulus of the magnetic moment, $M_s = |\mathbf{M}|$,
 and the approach to the Gibbs-Boltzmann equilibrium, see Eq.~(\ref{eq:equil-LGG}),
in the absence of non-conservative forces and time-dependent magnetic fields~\cite{Aron14}. We note that ${\rm D}_t^{(\alpha)}$ in the last term 
between the square brackets in Eq.~(\ref{eq:sLLG1-spintorque-adim}) can be simply replaced by ${\rm d}_t$ as the second term in Eq.~(\ref{eq:derivcovariante}) does not contribute 
due to the vector product with ${\mathbf M}$.

The chain rule for the time derivative of a function of the magnetisation vector governed by Eq.~(\ref{eq:sLLG1-spintorque-adim})
reads~\cite{Aron14} 
\begin{equation}
{\rm d}_t U({\mathbf M}) =
\frac{\partial U({\mathbf M})}{\partial M_i} {\rm d}_t M_i 
+ (1-2\alpha) 
 \frac{D \gamma_0^2}{1+\eta^2\gamma_0^2} 
P^\perp_{ij}
\frac{\partial^2 U({\mathbf M})}{\partial M_i \partial M_j}
\;,
\label{eq:chain rule-U}
\end{equation}
where $P^\perp_{ij} \equiv M_s^2 \delta_{ij} - M_i M_j$ is the projector on the plane perpendicular to $\mathbf{M}$.
A straightforward application  to $\mathbf{M}^2$ together with
 ${\mathbf M} \cdot {\rm D}_t^{(\alpha)} {\mathbf M}=0$
yields ${\rm d}_t \mathbf{M}^2=0$ and hence the conservation of the modulus.

The sLLG equation depends on several parameters.
$\gamma_0 = \gamma\mu_0$ is the product of 
$\gamma$, the gyromagnetic ratio relating the magnetisation
to the angular momentum, and $\mu_0$, the vacuum permeability constant.
The gyromagnetic factor is given by $\gamma=\mu_B g/\hbar$ (in our convention
$\gamma >0$
for the electronic spin) with $\mu_B$ Bohr's magneton and $g$ Lande's g-factor. 

The Gaussian white noise ${\mathbf H}(t)$ acts multiplicatively on the magnetisation. 
It has zero average and correlations characterised by the diffusion constant $D$,  
\begin{equation}
\langle H_{i}(t) \rangle_{\mathbf H} = 0
\;, \qquad
\langle  H_{i}(t)  H_{j}(t')  \rangle_{\mathbf H} = 2 D \ \delta_{ij}  \delta(t-t')
\; . 
\label{eq:noise-noise}
\end{equation}
We assume that the environment is in equilibrium at the temperature $k_BT\equiv\beta^{-1}$ 
yielding the Einstein relation 
\begin{equation}
D = \frac{\eta k_BT}{M_s V\mu_0}
\label{eq:D-def}
\end{equation}
 with $V$ the volume of the system and $\eta$ the friction coefficient that also appears in the last, dissipative, term in the rhs of Eq.~(\ref{eq:sLLG1-spintorque-adim}). Indeed, the term $-\eta {\rm D}^{(\alpha)}_t {\mathbf M}$ induces dissipation in the form introduced by Gilbert~\cite{Gilbert55}.

The deterministic magnetic field ${\mathbf H}_{\rm eff}$ collects conservative and 
non-con\-ser\-va\-tive contributions:
\begin{eqnarray}
{\mathbf H}_{\rm eff} = {\mathbf{H}}^{\rm c}_{\rm eff} + {\mathbf{H}}^{\rm nc}_{\rm eff}\,.
\end{eqnarray}
The former can be derived from a potential 
energy density $U$ as ${\mathbf{H}}^{\rm c}_{\rm eff} = -\mu_0^{-1} {\mathbf \nabla}_{\mathbf M} U$
whereas the latter does not admit such a representation.
$U$ can possibly have contributions from an externally applied magnetic field ${\mathbf H}_{\rm ext}$ (that we assume to be constant for simplicity) and from a 
local magnetic field typically generated by the anisotropy potential of the local crystal structure (the so-called crystal field)
\begin{equation}
U({\mathbf M}) = -\mu_0 {\mathbf M} \cdot {\mathbf H}_{\rm ext} + V_{\rm ani}({\mathbf M})
\; . 
\label{eq:equil-LGG}
\end{equation}
We shall only consider the case of time-reversal symmetric potentials, \textit{i.e.}
with the property $U(-{\mathbf M}, -{\mathbf H}_{\rm ext}) = U({\mathbf M}, {\mathbf H}_{\rm ext})$.

A timely example of a non-conservative ${\mathbf H}_{\rm eff}^{\rm nc}$ is the so-called spin-torque exchange. In the context of spintronics, the manipulation of the local magnetisation is performed by 
circulating a spin-polarised current of electrons through the ferromagnetic material. This  
can exchange angular momentum with the magnetisation \textit{via} the spin-torque term
\begin{equation}
{\mathbf H}^{\rm torq}_t = \chi_t \ \mathbf{M}_t \wedge \mathbf{p}_t 
\end{equation}
where   $\chi_t$  is a time-dependent parameter that is proportional to the externally controlled current 
$J_t$, and ${\mathbf p}_t$ is a unit vector indicating the spin polarisation of the 
incoming electrons that they typically acquire earlier by going through a thick layer 
of ferromagnetic material with a fixed magnetisation.
Dimensional analysis yields $[\chi]= [k_BT/(M_s^2 \mu_0 V)] = [H/M_s] = [(\gamma_0 t M_s)^{-1}]$.

Due to the fact that the magnetic fields appear under a vector product with  the magnetisation vector, 
only their projection on the perpendicular plane to ${\mathbf M}$ have an effect on the magnetic moment 
dynamics.

For ${\mathbf H}_{\rm eff}^{\rm nc}=0$ and $\mathbf{H}^{\rm c}_{\eff} = -\mu_0^{-1} \boldsymbol{\nabla}_\mathbf{M} U$,  
the dynamics approach the Gibbs-Boltzmann distribution 
\begin{equation}
P_{\rm GB}({\mathbf M}) = Z^{-1} \ \rme^{-\beta V U({\mathbf M})} 
\; . 
\end{equation}
Note that the partition function $Z= \int {\rm d}{\mathbf M} \ \rme^{-\beta V U({\mathbf M})}$ is a function of the inverse temperature, $\beta$, the external field, ${\mathbf H}_{\rm ext}$, and the parameters of the anisotropy potential $V_{\rm ani}$. One has $Z({\mathbf H}_{\rm ext}) = Z(-{\mathbf H}_{\rm ext})$.

From this stationary distribution, one simply shows the {\it static fluctuation-dissipation relation} 
between equilibrium susceptibility and magnetic fluctuation correlations:
\begin{equation}
\left.
\frac{\partial\langle M_i \rangle}{\partial {H_{\rm ext}}_j} \right|_{{\mathbf H}_{\rm ext}=0}
= \beta \mu_0 V \ \langle (M_i - \langle M_i\rangle) (M_j - \langle M_j\rangle ) \rangle
\; . 
\end{equation} 
We will prove the time-dependent fluctuation-dissipation theorem for this problem below (see also~\cite{Foros09}).

For simplicity, we study the dynamics for a system initially prepared in equilibrium 
with a Gibbs-Boltzmann distribution $P_\rmi\left(\mathbf{M}_{-\mathcal{T}}\right) 
= P_{\rm GB}\left(\mathbf{M}_{-\mathcal{T}}\right)$. In particular, we set $\chi_{-\mathcal{T}}= J_{-\mathcal{T}} = 0$.
In the absence of a drive, ${\mathbf H}_{\rm eff}^{\rm nc}=0$ $ \forall t$ and for a time-independent effective field ${\mathbf H}_{\rm eff}$, the system remains in thermal equilibrium. However, a finite drive ${\mathbf H}_{\rm eff}^{\rm nc} \neq 0$  or 
 a time-dependent effective field ${\mathbf H}_{\rm eff}(t)$
push the magnetic moment out of equilibrium.

\subsection{The path-integral formulation}

As shown in Ref.~\cite{Aron14}, the generating functional reads
\begin{equation}
{\cal Z}[\boldsymbol{\zeta}] =
\int \uD{[\mathbf{M}] \int \uD[\hat{\mathbf{M}}^\perp]}
\int \uD{[\hat{\mathbf{M}}^\parallel]}
\ \exp \left\{ S[\mathbf{M},\rmi\hat{\mathbf{M}}] +  \int \boldsymbol{\zeta}_t \cdot \mathbf{M}_t \right\}
\; .
\end{equation}
$\int \uD{[\mathbf{M}]}$ corresponds to integrating, at all times, 
over the vector field $\mathbf{M}$  on the $2$-sphere of radius $M_s$, 
 $\int \uD{[\hat{\mathbf{M}}^\perp]}$ corresponds to integrating over the auxiliary real vector field $\hat{\mathbf{M}}^\perp$ in the plane perpendicular  to $\mathbf{M}$. Correspondingly, $\int \uD{[\hat{\mathbf{M}}^\parallel]}$ represents the integration over the auxiliary real vector field in plane parallel to $\mathbf{M}$.
As ${\mathbf M} \cdot {\rm D}_t^{(\alpha)} {\mathbf M} =0 $, $\hat {\mathbf M}^\perp$ 
and ${\rm D}_t^{(\alpha)}{\mathbf M}$ are both perpendicular to ${\mathbf M}$ but not necessarily parallel.

The action functional can be expressed as the sum
\begin{align}
 S = S_{\rm det} + S_{\rm diss} + S_{\rm jac}
 \label{eq:total-action-Gilbert}
\end{align}
with
\begin{align}
 S_{\rm det} =&
 -\beta V U(\mathbf{M}_{-{\cal T}}, {\mathbf H}_{{\rm ext}_{-{\cal T}}}) 
 - 
 \ln Z
 ({\mathbf H}_{{\rm ext}_{-{\cal T}}})
+ \int \rmi \hat {\mathbf M}_t^\parallel \cdot {\rm D}_t^{(\alpha)} {\mathbf M}_t
 \nonumber \\
 &
  +   \int 
   \rmi\hat{\mathbf{M}}_t^\perp  \cdot  
   \left[  M_s^{-2} {\rm d}_t  \mathbf{M}_t \wedge \mathbf{M}_t 
   +  \gamma_0 {\mathbf{H}_{\eff}}_t 
   \right]
 \; , \label{eq:SLLGjac-det} 
 \\
 S_{\rm diss} =&
\;\;  \gamma_0
 \int
\rmi\hat{\mathbf{M}}_t^\perp  \cdot   \left[ D\gamma_0 \ \rmi\hat{\mathbf{M}}_t^\perp
  - \frac{\eta}{M_s} \ {\rm d}_t \mathbf{M}_t \right]
 \; . 
\label{eq:SLLGjac-diss}
\\
S_{\rm jac} =& 
 \;\;
 \frac{\alpha\gamma_0}{1+\eta^2\gamma_0^2} 
 \frac{1}{M_s}
 \int  \Big{[}
 M_s
 \epsilon_{ijk} {M_k}_t \partial_j {H^{\rm nc, \perp}_{{\eff}_{it}}}
- \eta\gamma_0
{P^\perp_{ij}}_t
\partial_j {H^\perp_{{\eff}_i}}_t
 \Big{]} 
\; .
 \label{eq:SLLGjac-jac}
 \end{align}
where $P^\perp_{ij} \equiv M_s^2 \delta_{ij} - M_i M_j$ is the projector on the plane perpendicular to $\mathbf{M}$
and $\partial_j$ is a short-hand notation for $\partial_{M_j}$. 
$S_{\rm det}$ encodes the initial conditions and the deterministic forces. $S_{\rm diss}$ encodes the hybridisation 
with the thermal bath.
The term in $\rmi \hat {\mathbf M}^\parallel_t$ imposes the conservation of the modulus of the magnetisation vector.
$S_{\rm jac}$ stems from the unicity of the solution to Eq.~(\ref{eq:sLLG1-spintorque-adim}) 
once the initial conditions and the noise history are specified. For magnetic fields that are independent
of the magnetisation vector, ${\mathbf H}_{\rm eff} = {\mathbf H}_{\rm ext}$,  $S_{\rm jac}$ vanishes.

The two-time correlation between any function of ${\mathbf M}$, $A(\mathbf{M})$, and, say, one component of the magnetisation, $M_j$, reads
\begin{align}
C_{A M_j}(t,t') =   \langle A({\mathbf M}(t))  M_j(t')  \rangle \;.
\end{align}
The linear response of the same observable $A$ measured at time $t$ to a previous perturbation $\widetilde {\mathbf H}$ that modifies the 
potential energy per unit volume $U$ according to $U \mapsto U - \mu_0^{-1} \ \widetilde {\mathbf H} \cdot {\mathbf M}$ (or equivalently ${\mathbf H}_{\rm eff} \mapsto {\mathbf H}_{\rm eff} + \widetilde {\mathbf H}$) is 
\begin{align}
R_{AM_j}(t,t') =  
\left. 
\frac{\delta \langle A({\mathbf M}(t))\rangle }{\delta {\widetilde{H}_j(t')}}
\right|_{\widetilde{\mathbf H}={\mathbf 0}} 
&
=  \ \langle A({\mathbf M}(t))  \gamma_0 \rmi\hat M^\perp_j(t')  \rangle
\;.
\label{eq:linear-response-M}
\end{align}

\subsection{Fluctuations}

Equilibrium dynamics are ensured as long as the system is initially prepared in equilibrium at a 
given temperature $T$ and under conservative forces, and that it is subsequently evolved under the same time-independent potential 
forces and in contact with an environment at the same temperature. In our setup, this corresponds to setting
the drive to zero, {\it i.e.} ${\mathbf H}_{\rm eff}^{\rm nc} = 0$ at all times, and taking ${\mathbf H}_{\rm ext}$ constant 
and $V_{\rm ani}$ not explicitly dependent on time.

\subsubsection{The time-reversal transformation}
 One can prove that the action in 
Eqs.~(\ref{eq:total-action-Gilbert})-(\ref{eq:SLLGjac-jac}), and more generally the 
full generating functional, are invariant under the following variable and discretisation parameter transformations
\begin{eqnarray}
{\cal T}_{\rm eq} = \left\{
\begin{array}{rcl}
\mathbf{M}_t &\mapsto& - \mathbf{M}_{-t} \; , \\
\gamma_0 \ \rmi\hat{\mathbf{M}}^\perp_t &\mapsto&  
 - \gamma_0 \ \rmi{\hat{\mathbf{M}}}^\perp_{-t} - \beta V\mu_0 \
{\rm d}_t \mathbf{M}_{-t}
\; , \\
\rmi \hat {\mathbf M}^\parallel_t &\mapsto& \rmi \hat {\mathbf M}^\parallel_{-t}
\; , 
\end{array}
\right.
\label{eq:symmetry-M}
\end{eqnarray}
if one simultaneously changes the discretisation parameter
\begin{equation}
\alpha \mapsto  \overline{\alpha} \equiv 1-\alpha
\end{equation}
and simultaneously reverses all external constant magnetic fields
\begin{equation}
{{\mathbf H}_{\rm ext}} \mapsto \overline{{\mathbf H}}_{\rm ext} \equiv  - {\mathbf H}_{\rm ext}
\; . 
\end{equation}

This discrete symmetry of the dynamical action encodes all the features of equilibrium dynamics and it is broken out of equilibrium.

\paragraph{Case without anisotropy potential, $V_{\rm ani}=0$.}
Let us start the proof by treating the simpler case in which  ${\mathbf H}_{\rm eff}={\mathbf H}_{\rm ext}$. It is easy to see, similarly to the one-dimensional example of Sec.~\ref{sec:one-dim}, that the dissipative terms in $S_{\rm diss}$ are invariant independently of the other terms of the action functional. We do not reproduce here this calculation as it is quite straightforward. We simply mention that we do not need to use $\alpha \mapsto 1-\alpha$ since in this case $S_{\rm diss}$ does not depend on $\alpha$ explicitly.
Let us now discuss the invariance of $S_{\rm det}$ and $S_{\rm jac}$. Since ${\mathbf H}_{\rm eff}={\mathbf H}_{\rm ext}$, $S_{\rm jac}$ vanishes and is therefore trivially invariant under ${\cal T}_{\rm eq}$.
For $S_{\rm det}$, we look separately at the terms proportional to $\rmi \hat {\mathbf M}_t^\perp$ and $\rmi \hat {\mathbf M}_t^\parallel$.
For the former
\begin{eqnarray*}
&& S^\perp_{\rm det}[{\cal T}_{\rm eq} \rmi \hat {\mathbf M}, {\cal T}_{\rm eq} {\mathbf M}, - {\mathbf H}_{\rm ext}] 
=
\beta V \mu_0 \ {\mathbf M}_{{\cal T}} \cdot {\mathbf H}_{\rm ext} - \ln Z(-{\mathbf H}_{\rm ext})
\nonumber\\
&&
\qquad\quad
+
\int [-\rmi \hat {\mathbf M}^\perp_{-t} - \beta V \mu_0 \gamma^{-1}_0 \ {\rm d}_t {\mathbf M}_{-t}] 
\cdot 
[
M_s^{-2} {\rm d}_t {\mathbf M}_{-t} \wedge {\mathbf M}_{-t} - \gamma_0 {\mathbf H}_{\rm ext} 
]
\; . 
\end{eqnarray*}
Given the symmetry property $U(-{\mathbf M}, -{\mathbf H}_{\rm ext}) = U({\mathbf M},{\mathbf H}_{\rm ext})$, we have  \linebreak
$Z(-{\mathbf H}_{\rm ext}) = Z({\mathbf H}_{\rm ext})$.
We first change $t \mapsto -t$ as the integration variable in the temporal integrals and we next rearrange terms to write
\begin{eqnarray}
&& S^\perp_{\rm det}[{\cal T}_{\rm eq} \rmi \hat {\mathbf M}, {\cal T}_{\rm eq} {\mathbf M}, - {\mathbf H}_{\rm ext}] 
=
\beta V \mu_0 \ {\mathbf M}_{{\cal T}} \cdot {\mathbf H}_{\rm ext} - \ln Z({\mathbf H}_{\rm ext})
\nonumber\\
&&
\qquad\qquad
+
\int [\rmi \hat {\mathbf M}^\perp_{t} - \beta V \mu_0 \gamma^{-1}_0 {\rm d}_t {\mathbf M}_{t}] 
\cdot 
[
M_s^{-2} {\rm d}_t {\mathbf M}_{t} \wedge {\mathbf M}_{t} + \gamma_0 {\mathbf H}_{\rm ext} 
]
\nonumber\\
&& 
\quad
=
\beta V \mu_0 \ {\mathbf M}_{{\cal T}} \cdot {\mathbf H}_{\rm ext} - \ln Z({\mathbf H}_{\rm ext})
+
\int \rmi \hat {\mathbf M}^\perp_{t} 
\cdot 
[
M_s^{-2} {\rm d}_t {\mathbf M}_{t} \wedge {\mathbf M}_{t} + \gamma_0 {\mathbf H}_{\rm ext} 
]
\nonumber\\
&&
\qquad\qquad
-  \beta V \mu_0  \int {\rm d}_t {\mathbf M}_{t} \cdot {\mathbf H}_{\rm ext}
\; . 
\label{eq:Sdetutil}
\end{eqnarray}
The first integral has the original form. The second integral can be computed directly, as ${\mathbf H}_{\rm ext}$ 
is independent of ${\mathbf M}$. One recovers the boundary terms at ${\cal T}$ and $-{\cal T}$; 
one cancels the first term in the rhs, the other one builds the initial probability weight.

The term that imposes the spherical constraint is invariant on its own if we use
\begin{equation}
 \alpha \mapsto 1-\alpha
 \qquad \mbox{and} \qquad 
 \rmi \hat {\mathbf M}^\parallel_t \mapsto  \rmi {\hat {\mathbf M}}^\parallel_{-t} 
 \end{equation}
 Indeed, ${\rm D}_t^{(\alpha)} \mapsto {\rm D}_{t}^{(1-\alpha)} = - {\rm D}_{-t}^{(\alpha)}$ and 
\begin{eqnarray}
\int \rmi \hat {\mathbf M}^\parallel_t \cdot {\rm D}_t^{(\alpha)} {\mathbf M}_t 
\mapsto 
\int \rmi \hat {\mathbf M}^\parallel_{-t} \cdot (-{\rm D}_{-t}^{(\alpha)}) (- {\mathbf M}_{-t} )
=
\int \rmi \hat {\mathbf M}^\parallel_{t} \cdot {\rm D}_{t}^{(\alpha)} {\mathbf M}_{t} 
\; . 
\end{eqnarray}
Note the different transformation rules on $\rmi \hat {\mathbf M}^\perp$ and $\rmi {\mathbf M}^\parallel$. 
The transformation $\alpha \mapsto 1-\alpha$ is needed so that the two terms in ${\rm D}_t^{(\alpha)}$ 
be odd under time reversal and ${\rm D}_t^{(\alpha)}$ behave as a usual time derivative.

\paragraph{Case with an anisotropy potential, $V_{\rm ani}\neq 0$.}
Let us now examine the generic case in which there is an anisotropy potential and $S_{\rm jac}$ no longer vanishes.
The analysis of the dissipative part of the action is identical to the one we discussed in the previous
subsection. As for the combined contributions $S_{\rm det}+S_{\rm jac}$, we proceed as follows.
We start from Eq.~(\ref{eq:Sdetutil}) conveniently generalised to take into account the fact that 
${\mathbf H}^{\rm c}_{\rm eff} = - \mu_0^{-1} {\mathbf \nabla}_{\mathbf M} U$:
\begin{eqnarray}
&& S^\perp_{\rm det}[{\cal T}_{\rm eq} \rmi \hat {\mathbf M}, {\cal T}_{\rm eq} {\mathbf M}, - {\mathbf H}_{\rm ext}] 
=
- \beta V U(-{\mathbf M}_{{\cal T}}, -{\mathbf H}_{\rm ext}) - \ln Z({\mathbf H}_{\rm ext})
\nonumber\\
&&
\qquad\qquad
+
\int \rmi \hat {\mathbf M}^\perp_{t} 
\cdot 
[
M_s^{-2} {\rm d}_t {\mathbf M}_{t} \wedge {\mathbf M}_{t} + \gamma_0 {\mathbf H}_{{\rm eff}_t}
]
\nonumber\\
&&
\qquad\qquad
+ \ \beta V  \int {\rm d}_t {\mathbf M}_{t} \cdot {\mathbf \nabla}_{{\mathbf M}_t} U({\mathbf M}_t, {\mathbf H}_{\rm ext})
\; . 
\label{eq:Sdetutil2}
\end{eqnarray}
We use now Eq.~(\ref{eq:chain rule-U})
\begin{eqnarray}
{\rm d}_t {\mathbf M} \cdot {\mathbf \nabla}_{\mathbf M} U = {\rm d}_t M_i  \ \partial_i U = 
 {\rm d}_t U - D(1-2\alpha) \frac{\gamma_0^2}{1+\eta^2\gamma_0^2} P_{ij}^\perp \partial_i \partial_j U
\end{eqnarray}
 in the last term and we obtain $+\beta V U({\mathbf M}_{\cal T}, {\mathbf H}_{\rm ext}) -
 \beta V U({\mathbf M}_{-{\cal T}}, {\mathbf H}_{\rm ext}) $
 after integrating the total time derivative.
Using the property $  U(-{\mathbf M}_{\cal T}, -{\mathbf H}_{\rm ext}) = U({\mathbf M}_{\cal T}, {\mathbf H}_{\rm ext}) $, the first term cancels the first term in the rhs in (\ref{eq:Sdetutil2})  and the second term reconstructs the exponential weight in the initial distribution.
We therefore have 
\begin{eqnarray}
&& S^\perp_{\rm det}[{\cal T}_{\rm eq} \rmi \hat {\mathbf M}, {\cal T}_{\rm eq} {\mathbf M}, - {\mathbf H}_{\rm ext}] 
 = 
S^\perp_{\rm det}[ \rmi \hat {\mathbf M},  {\mathbf M}, {\mathbf H}_{\rm ext}] 
\nonumber\\
&&
\qquad\qquad\qquad\qquad
-  \frac{ (1-2\alpha)\eta\gamma_0^2}{1+\eta^2\gamma_0^2} \frac{1}{M_s\mu_0} \int P_{ij}^\perp \partial_i \partial_j U
\; , 
\label{eq:Sdetperp}
\end{eqnarray}
where we replaced $D$ by its definition in Eq.~(\ref{eq:D-def}), while
\begin{equation}
S_{\rm jac}[{\cal T}_{\rm eq} {\mathbf M}, 1-\alpha] 
= 
 \frac{[\alpha + (1-2\alpha)]\eta \gamma^2_0}{1+\eta^2\gamma_0^2} \frac{1}{M_s\mu_0} \int P_{ij}^\perp \partial_i \partial_j U
 \; . 
\end{equation}
We notice that the first term is what we need to build $S_{\rm jac}[{\mathbf M}, \alpha]$ and the 
last term cancels the remaining one in Eq.~(\ref{eq:Sdetperp}). 

The invariance of the term imposing the spherical constraint works in the same way as in the $V_{\rm ani}=0$ case.

We have therefore completed the proof of invariance of the action under the transformation in (\ref{eq:symmetry-M}). 

\subsubsection{The fluctuation-dissipation theorem}

Applying this symmetry to the linear response, see Eq.~(\ref{eq:linear-response-M})
we obtain
\begin{align}
& R_{A{\mathbf M}_j}(t,t') \equiv \langle A(t) \, \gamma_0 \rmi \hat M^\perp_j (t') \rangle
\nonumber\\
& \quad = -\langle A_\rmr(-t) \, \gamma_0 \rmi \hat{M}^\perp_j(-t') \rangle 
- \beta V \mu_0 \  \langle A_\rmr(-t) {\rm d}_{t'} M_j(-t') \rangle \nonumber \\
& \quad = -R_{A_\rmr {\mathbf M}_j}(-t,-t') - \beta V \mu_0  \ {\rm d}_{t'} C_{A_r {\mathbf M}_j}(-t,-t')
\nonumber\\
& \quad = -R_{A_\rmr {\mathbf M}_j}(-t,-t') + \beta V \mu_0  \  {\rm d}_{t'} C_{A {\mathbf M}_j}(t,t')
\;,
\end{align}
where $A_\rmr$  is the time-reversed observable of $A$. In the last step, we applied the transformation once more to the last term.
All averages are taken with the unperturbed
action measured in the original variables and with the $\alpha$ parameter, what we would call 
$S[{\mathbf M}, \rmi \hat {\mathbf M}; \alpha]$.
Using the causality of the response, $R_{A_r M_j}(-t,-t')=0$ for $t>t'$ and the time-translational invariance of  equilibrium dynamics, one may simplify the expression above to
\begin{align}
R_{A{\mathbf M}_j}(\tau) &= - \beta V \mu_0 \, \Theta(\tau) \ {\rm d}_{\tau}C_{A{\mathbf M}_j} (\tau)  
\end{align}
where we introduced $\tau = t - t'$ and $\Theta(\tau)$ is the Heaviside step function. Note that this relation applies to any observable $A$. Higher order fluctuation-dissipation 
relations of this kind can by easily derived.

\subsubsection{Broken symmetry and fluctuation theorems}

In this three-dimensional vector problem, the spin-torque due to a spin-polarised current $\chi_t \neq 0$ for $t>-{\cal T}$ is a non-conservative force that drives the magnetic moment out of equilibrium.  
It gives rise to the following contribution to the deterministic action, 
\begin{equation}
S_{\rm det}^{\rm torq}
= \gamma_0
\int \rmi {\hat {\mathbf M}}_t^\perp \cdot \chi_t \ ({\mathbf M}_t \wedge {\mathbf p}_t )
\; , 
\end{equation}
and to an extra term in the Jacobian, 
\begin{align}
S_{\rm jac}^{\rm torq} =
 \frac{2\alpha\gamma_0}{1+\eta^2\gamma_0^2} \int  \chi_t \ \mathbf{M}_t  \cdot \mathbf{p}_t 
 \; . 
\end{align}
When evaluated with the time-reversed variables
\begin{equation}
\overline \chi_t = - \chi_{-t} \qquad\qquad \overline {\mathbf p}_t = - {\mathbf p}_{-t}
\; , 
\end{equation}
(ensuring notably that ${\mathbf H}^{\rm torq}$ is odd under time reversal), the deterministic part of the action functional corresponding to the time-reversed dynamics reads
\begin{eqnarray}
&& S_{\rm det}^{\rm torq}[{\cal T}_{\rm eq} {\mathbf M}, {\cal T}_{\rm eq} \rmi \hat {\mathbf M}; \overline \chi, \overline {\mathbf p}]
=
\int [ - \gamma_0 \rmi {\hat {\mathbf M}}_{-t}^\perp - \beta V \mu_0 {\rm d}_t {\mathbf M}_{-t} ] 
\cdot \chi_{-t} \ (-{\mathbf M}_{-t} \wedge {\mathbf p}_{-t} )
\nonumber\\
&&
\qquad\qquad
=
 \int [ \gamma_0 \rmi {\hat {\mathbf M}}_{t}^\perp \cdot \chi_{t} \ ({\mathbf M}_{t} \wedge {\mathbf p}_t )
+\beta V \mu_0 {\rm d}_{-t}{\mathbf M}_{t} 
\cdot \chi_{t} \ ({\mathbf M}_{t} \wedge {\mathbf p}_t )
]
\nonumber\\
&&
\qquad\qquad
=
S_{\rm det}^{\rm torq}[{\mathbf M}, \rmi \hat {\mathbf M};  \chi, {\mathbf p}]
-\beta V \mu_0\int {\rm d}_{t} {\mathbf M}_{t} 
\cdot \chi_{t} \ ({\mathbf M}_{t} \wedge {\mathbf p}_t )
\; . 
\end{eqnarray}
We see that the term that is generated cannot be partially integrated away, as in the potential 
case, since ${\mathbf H}^{\rm torq}$ is not the gradient of a potential.
The spin-torque contribution to the Jacobian transforms as
\begin{align}
 S_{\rm jac}^{\rm torq}[{\cal T}_{\rm eq} {\mathbf M}; \overline \alpha, \overline \chi, \overline {\mathbf p}] 
& = -\frac{2 (1-\alpha) \gamma_0}{1+\eta^2\gamma_0^2} 
\int  \chi_t \ \mathbf{M}_t  \cdot \mathbf{p}_t 
\nonumber\\
&  = S_{\rm jac}^{\rm torq}[{\mathbf M}; \alpha, \chi, {\mathbf p} ] 
+ \Delta S_{\rm jac}^{\rm torq}[{\mathbf M};  \chi, {\mathbf p} ] 
\end{align}
with 
\begin{align}
 \Delta S_{ \rm jac}^{\rm torq}[{\mathbf M}; \chi, {\mathbf p} ] 
& = -\frac{2\gamma_0}{1+\eta^2\gamma_0^2} 
\;\int  \chi_t \ \mathbf{M}_t  \cdot \mathbf{p}_t 
\; . 
\end{align}
The rest of the action remains invariant under this 
transformation as it was in the absence of the spin-torque term and with no time-dependent 
parameter dependencies in the effective field ${\mathbf H}_{\rm ext}$.
Ultimately, we obtain
\begin{align}
\Delta S \equiv \Delta S_{\rm jac}^{\rm torq}[{\mathbf M};  \chi, {\mathbf p} ]  - \beta V \mu_0 W[{\mathbf M}; \chi, {\mathbf p}]\;.
\end{align}
with 
\begin{equation}
W[{\mathbf M}; \chi, {\mathbf p}] \equiv  \int \chi_t \ \mathbf{p}_t \cdot \left(  {\rm d}_t \mathbf{M}_t \wedge \mathbf{M}_t \right)
\end{equation}
the work performed by the spin-torque term.
This result implies the relation
\begin{align}
\frac{P_{\rm B}[\mathbf{M},  \rmi \hat {\mathbf M}]}{P_{\rm F}[\mathbf{M}, \rmi \hat {\mathbf M}]}  = \rme^{\Delta S} \label{eq:pdf-identity}
\end{align}
where we defined the forward and backward path probability distributions
\begin{align}
& P_{\rm F}[\mathbf{M}, \rmi \hat {\mathbf M}] \equiv   P[\mathbf{M},  \rmi \hat {\mathbf M}; \alpha, \mathbf{H}_{\rm ext}, \chi, \mathbf{p}]\;, \label{eq:forwLLG}\\
& P_{\rm B}[\mathbf{M}, \rmi \hat {\mathbf M}] \equiv   P[{\cal T}_{\rm eq} \mathbf{M}, {\cal T}_{\rm eq} \rmi \hat {\mathbf M}; \overline{\alpha}, \overline{\mathbf{H}}_{\rm ext}, \overline{\chi}, \overline{\mathbf{p}}] \;.\label{eq:backLLG}
\end{align}
Multiplying the identity (\ref{eq:pdf-identity}) by  a generic observable $A$ which can depends on $\mathbf{M}$, $\rmi \hat{\mathbf{M}}$ and $\mathbf{H}_{\rm ext}$, one obtains
\begin{align}
\langle A(\mathbf{M},\rmi \hat{\mathbf{M}},\mathbf{H}_{\rm ext}) \rangle_{\rm B} = 
\langle A(\mathbf{M},\rmi \hat{\mathbf{M}},\mathbf{H}_{\rm ext}) \ \rme^{\Delta S} \rangle_{\rm F}
\end{align}
where the subscripts ${\rm F}$ and ${\rm B}$ stand for averaging with the forward and backward path probability distributions defined in 
Eqs.~(\ref{eq:forwLLG}) and (\ref{eq:backLLG}), respectively.
In particular, for $A = 1$, this boils down to the Jarzynski equality~\cite{Jarzynski97,Jarzynski04} that reads in this case
\begin{align}
\langle  \rme^{\Delta S_{\rm jac}^{\rm torq}[{\mathbf M};  \chi, {\mathbf p} ]  -\beta V \mu_0 W[{\mathbf M}; \chi, {\mathbf p}]} \rangle_{\rm F}= 1\;.
\end{align}

\section{Conclusions and outlook}
\label{sec:conclusions}

In this paper, we studied Markov stochastic processes with multiplicative white noise and adequately drifted to ensure their approach to the usual Gibbs-Boltzmann measure under equilibrium conditions. In this respect, our viewpoint is different from the one in Ref.~\cite{arenas2012} where no drift force was added to the stochastic differential equation and the dynamics approached a non-Gibbs-Boltzmann stationary state.

In recent years, thermodynamic relations and concepts in out-of-equi\-lib\-rium stochastic processes have been searched for. 
{In this paper, we proposed a field-theoretical derivation of fluctuation theorems for Markov stochastic processes with multiplicative white noise. This approach, based on a particular symmetry breaking of the path integral representation of the generating functional, extends our previous work~\cite{AronLeticia2010} by showing that a single model-independent field transformation, $\mathcal{T}_{\mathrm{eq}}$, can generate all the fluctuation theorems for any stochastic evolution, in the presence of white or colored, additive or multiplicative noise.}

One could ask whether the effective temperature idea~\cite{cugl:review} applies to multiplicative-noise processes that are not able to reach equilibrium with their surroundings and whether once set out of equilibrium they would satisfy fluctuation theorems in the way discussed in Ref.~\cite{Zamponi05} for additive-noise processes.
Exploring the {\it stochastic thermodynamics} and energetics~\cite{seifert2008,Sekimoto98} proposals for multiplicative-noise Markov processes should also be an interesting research project.


\appendix

\section{Stochastic calculus}
\label{subsec:chain rule}

\subsection{Chain rule}

We examine the time derivative of a generic function, $[F(x_{n+1}) - F(x_n)]/{\rmd t}$,  with the stochastic variable $x$ 
governed by the Langevin equation with multiplicative white noise,
\begin{equation} \label{eq:xeom}
x_{n+1} - x_n = f(\overline x_n) {\rmd t} + g(\overline x_n) {\rmd} W_n
\; , 
\end{equation}
with $(\rmd W_n)^2 \simeq 2D \ {\rmd t} $.

If we expand $x_{n+1}$ in $F(x_{n+1})$, and $x_n$ in $F(x_n)$, around the generic $\alpha$ point 
$\bar x_n = \alpha x_{n+1} + (1-\alpha) x_n$ we
obtain
\begin{align}
& F(x_{n+1}) - F(x_n) = F(\bar x_n + (1-\alpha)(x_{n+1} - x_n)) - F(\bar x_n - \alpha(x_{n+1} - x_n)) \nonumber \\
& 
\qquad
= (x_{n+1} - x_n) F'(\bar x_n) + \frac{1}{2}(1-2\alpha) (x_{n+1} - x_n)^2 F''(\bar x_n) + \mathcal{O}(\rmd x^3)
\nonumber
\end{align}
with $\rmd x= x_{n+1} - x_n$. 
Using now Eq.~(\ref{eq:xeom}) to replace $(x_{n+1} - x_n)^2 $ by  $2 D g(\bar x_n)^2 {\rmd t} + \mathcal{O}({\rmd t}^{3/2})$, 
\begin{align}
F(x_{n+1}) - F(x_n) &= (x_{n+1} - x_n)  F'(\bar x_n) + (1-2\alpha)D g(\bar x_n)^2 F''(\bar x_n) {\rmd t} + \mathcal{O}(\rmd x^3)
\nonumber
\end{align}
that in the limit ${\rmd t} \to 0$ becomes
\begin{equation}
\frac{F(x_{n+1}) - F(x_n)}{{\rmd t}}= \frac{x_{n+1} - x_n}{{\rmd t}}  F'(\bar x_n) + (1-2\alpha)D g(\bar x_n)^2 F''(\bar x_n) 
\; . 
\label{eq:chain ruledisc} 
\end{equation} 
This expression is written as the generalised chain rule~\cite{gardiner,vanKampen}
\begin{align}
\rmd_t F(x)  &= \rmd_t x \ F'(x) + (1-2\alpha) D g^2(x) F''(x)
\; . 
\end{align}

\subsection{From the $\alpha$ to the  Stratonovich prescription}

One can transform a stochastic equation in the generic $\alpha$-prescription into one in the Stratonovich mid-point prescription by simply expanding
the arguments of $f$ and $g$ around the latter points. More precisely, let us start from an equation in the $\alpha$-prescription 
\begin{equation}
\frac{{\rm d}x}{{\rm d} t} = f(x) + g(x) \xi \qquad \Leftrightarrow \qquad
x_{n+1} - x_n = f(\overline x_n) {\rm d}t + g(\overline x_n) {\rm d}W_n 
\; , 
\label{eq:alpha-eq}
\end{equation}
{\it i.e.} $\overline x_n = x_n + \alpha (x_{n+1} - x_n)$.
The Stratonovich mid-points are $x^S_n = x_n + \frac{1}{2} (x_{n+1} - x_n)$, and
$\alpha$ and $S$ points are related by 
\begin{equation}
\overline x_n = x_n^S - \frac{1}{2} (1-2\alpha) (x_{n+1} - x_n)
\; . 
\end{equation}
Expanding now Eq.~(\ref{eq:alpha-eq}) around the $S$ mid-points one finds
\begin{equation}
x_{n+1} - x_n = [f(x_n^S) - (1-2\alpha) D g(x_n^S) g'(x_n^S) ]  {\rm d}t + g(x_n^S) {\rm d} W_n 
\end{equation}
where we dropped contributions of ${\cal O}({\rm d}t^{3/2})$ and we used $({\rm d}W_n)^2 = 2D{\rm d}t$.
The function $g$ that multiplies ${\rm d}W_n$ is evaluated now at the $S$ point $x_n^S$ and in this sense this is 
an equation in the Stratonovich prescription.

This same strategy can be followed to transform an equation from the $\alpha$ to the $\alpha'$-prescription at the price of
modifying the force with an adequate drift term.

\section{From multiplicative to additive noise}
\label{app:mult-addit}

It is often found in the literature that a multiplicative-noise process can be mapped to an additive-noise process, and that in the latter formulation all subtleties linked to the discretisation prescription can be simply forgotten. Here we show that, while indeed such a mapping exists~\cite{Schenzle79}, the discretisation used to define the original multiplicative-noise process enters the additive-noise process in the form of a non-trivial drift force.

 \label{sec:mapping}
Let us re-parametrise the original equation of motion, Eq.~(\ref{eq:x-eom}) with $g(x)\equiv 1/k(x)$, such that the origin of the multiplicative noise and of the velocity in the lhs can be tracked back to a non-linear coupling to a thermal bath of oscillators, see the discussion below Eq.~(\ref{eq:eq-a-la-Aron-Biroli-LFC}). We write it as the equation (in the $\alpha$-prescription)
\begin{align}\label{eq:eom3b-St}
 k^2(x) \rmd_t x(t) &= f(x) + k(x) \xi(t) 
 \; , 
\end{align}
where we also re-parametrized $f$ by $f(x) k^2(x) \mapsto f(x)$ such that $f$ can now be thought of as a true force, in units of Newtons, possibly deriving from a potential $f(x) = - V'(x)$. If we now divide this equation by $k(x)$, we obtain an equation in which the noise appears additively.
However, it has to be treated with great care because the term $k(x) \rmd_t x$ hides subtleties associated with the discretisation.

\comments{

\textcolor{red}{We can simply erase the Strato case, I keep it for the moment}

\textcolor{gray}{
\paragraph{Stratonovich prescription.}
\label{subsubsec:strato}
We now divide Eq.~(\ref{eq:eom3b-St}) by $k$ 
and we introduce $K$, the primitive of $k$, such that $K'=k$. We then have
\begin{align}\label{eq:eom4-strato}
K'(x) \rmd_t x  &= f_k(x)/K'(x) + \xi(t) 
\; . 
\end{align}
We set the prescription to the Stratonovich  one, $\alpha=1/2$, so that the conventional calculus applies.
We perform the following change of variable change variables from $x$ to $\tilde u$ according to  the non-linear transformation
\begin{align}
\tilde u \equiv K(x)
\; . 
\end{align}
As the usual rules of calculus apply we obtain,
\begin{align}\label{eq:eom5-strato}
\rmd_t \tilde{u}  &= \tilde{f}_k(\tilde u) + \xi(t) 
\; , 
\end{align}
where we introduced $\tilde f_k(\tilde{u}) \equiv f_k(x(\tilde u)) / K'(x(\tilde u))$ \textit{i.e.} $\tilde f_k(\tilde u) = f_k(x(\tilde u)/k(x(\tilde u))$. 
Equation~(\ref{eq:eom5-strato}) is a Stratonovich Langevin equation with additive noise. 
}
\paragraph{The $\alpha$-prescription.}
\label{subsubsec:alpha}

}

Equation~(\ref{eq:eom3b-St}) is defined in the $\alpha$-prescription, for which unusual rules of calculus apply. Re-writing it momentarily as
\begin{align}\label{eq:eom5}
\rmd_t x &= f(x)/k^2(x) + 1/k(x) \ \xi(t) 
\;,
\end{align}
one can show that the associated chain rule reads [see the proof in Gardiner's book~\cite{gardiner} for $\alpha=0$ or $1/2$ and recalled in Eq.~(\ref{eq:x-chain}) for any $\alpha$] 
\begin{equation}
\rmd_t K(x) = K'(x) \rmd_t x + D (1-2\alpha) \ K''(x)/K'(x)^2
\label{eq:chain-K}
\; . 
\end{equation}
where we introduced $K$ such that $K'(x)\equiv k(x)$.
Using Eq.~(\ref{eq:chain-K}),  Eq.~(\ref{eq:eom5}) can be re-written as
\begin{equation}
\rmd_t K - D(1-2\alpha) K''(x)/K'(x)^2 = f(x)/K'(x) +\xi(t)\;.
\end{equation}
Let us now perform the change of variable from $x$ to $u$ according to  the non-linear transformation
\begin{align}
u \equiv K(x)
\end{align}
to get the additive-noise process
\begin{align}\label{eq:eom5-alpha}
\rmd_t u  &= D(1-2\alpha) K''(x(u))/K'(x(u))^2  +\tilde f(u) +\xi(t)
\end{align}
where we introduced $\tilde f(u) \equiv f(x(u)) / K'(x(u))$. 
We finished mapping the original multiplicative-noise process in Eq.~(\ref{eq:eom3b-St}) to an additive-noise Langevin equation with a simple time-dervivative $\rmd_t u$ in the lhs, but the first term in the rhs is non-trivially inherited from the discretisation of the original multiplicative-noise process and depends explicitly on $\alpha$ and $k$. It only vanishes for $\alpha=1/2$ (Stratonovich calculus) or $k'=0$ (additive noise in the original Langevin equation).

If we re-parametrize back to the original notations of Eq.~(\ref{eq:x-eom}), \textit{i.e.} $k \mapsto 1/g$ and $f \mapsto f/g^2$, we obtain the mapping of Eq.~(\ref{eq:x-eom}) to
\begin{align}
\rmd_t u  &=  - D(1-2\alpha) g'(x(u)) + f(x(u))/g(x(u))+\xi(t)
\; . 
\label{eq:naive-eq-add-noise}
\end{align}

\section{Stationary distribution and drift term}
\label{subsec:stat-Langevin}

We revisit here the need for  a drift term to ensure the approach to the Gibbs-Boltzmann distribution in the asymptotic long-time limit by working with the Langevin equation approach exclusively. We show that, given the generic multiplicative-noise equation~(\ref{eq:eom3b-St}), the stationary distribution is not of Boltzmann form unless a force is added to the conservative force, consistently with what we found with the Fokker-Planck approach in Sec.~\ref{subsec:FP}. 

\paragraph{Case of a Stratonovich prescription.}
Let us first treat the simpler case of $\alpha = 1/2$. 
We learned above that Eq.~(\ref{eq:eom3b-St}) can be re-written as the additive-noise equation of motion
\begin{align}
\label{eq:eom5s}
\rmd_t u  &= \tilde{f}(u) + \xi(t) 
\end{align}
with 
\begin{align}
u \equiv K(x), \quad K'(x) \equiv k(x), \quad \tilde f(u) \equiv f(x)/k(x)
\; . 
\end{align}
The stationary probability distribution of the stochastic variable $ u$ governed by Eq.~(\ref{eq:eom5s}) is 
\begin{align}
\widetilde{P}_{\rm st}(u) =N  \rme^{ - \beta \widetilde{V}(u) }
\qquad 
\mbox{with}
\qquad
 \widetilde{V}(u)  \equiv -\int^{u} \ud{u'} \tilde{f}(u')
\end{align}
 and $N$ a normalisation constant. Switching back to the stochastic variable $x$ governed by 
Eq.~(\ref{eq:eom5}), the corresponding probability distribution can be recovered as
\begin{align}
P_{\rm st}(x) & = \left| \frac{\rmd u(x)}{\rmd x} \right| \widetilde{P}_{\rm st}(u(x)) 
= N \left| k(x)\right| \ \rme^{ \beta  \int^{u(x)} \ud{u'} \tilde{f}(u') }
\nonumber\\
& = N \left| k(x)\right| \ \rme^{\beta  \int^{x} \ud{x'} f(x')  }
\; . 
\end{align}
This corresponds to the usual Gibbs-Boltzmann distribution if we allow ourselves to work with
\begin{align}
 f = -  V'- k_B T\,  k'/k
\; . 
\end{align} 
With this choice, the equation of motion becomes
\begin{align}\label{eq:eom3d}
 k^2(x) \rmd_t x(t) &= - V'(x) - k_BT k'(x)/k(x) + k(x) \xi(t) 
 \; . 
\end{align}
If we re-parametrize back to the original notations of Eq.~(\ref{eq:x-eom}), \textit{i.e.} $k \mapsto 1/g$ and $f \mapsto f/g^2$, we obtain
\begin{equation}
\rmd_t x(t) = - g^2(x) V'(x) + k_BT g'(x) g(x) + g(x) \xi(t) 
\end{equation}
which is Eq.~(\ref{eq:x-eom-drifted}) in the case $\alpha=1/2$.

\paragraph{Generic $\alpha$-prescription.}

The generalisation to the generic $\alpha$ case is straightforward.
Equation~(\ref{eq:eom3b-St}) can be re-written as Eq.~(\ref{eq:eom5-alpha})
with a stationary probability distribution 
$
\widetilde{P}_{\rm st}(u) = N \rme^{ - \beta \widetilde{V}_k(u) }
$
for $u$ and 
\begin{equation}
 \widetilde{V}(u)  \equiv -\int^{u} \ud{u'} [\tilde{f}(u') + D(1-2\alpha) K''(x(u'))/K'(x(u'))^2 ]
 \; . 
 \end{equation} 
Returning to the the stochastic variable $x$, \textit{via} $u = K(x)$, the corresponding stationary probability distribution is 
\begin{align}
P_{\rm st}(x) 
= N  |k(x)| \rme^{ (1-2\alpha) \ln k(x) + \beta \int^x \ud{x'} f(x') }
\; . 
\label{eq:stat-dist-alpha}
\end{align}
This corresponds to the usual Gibbs-Boltzmann distribution if we  work with
\begin{align}
f(x) = - V'(x) - 2k_B T (1-\alpha) k'(x)/k(x)
\; . 
\end{align} 
The Langevin equation becomes
\begin{align}\label{eq:eom3d-alpha}
 k^2(x) \rmd_t x(t) &= -V'(x) - 2k_BT (1-\alpha) \ k'(x)/k(x) + k(x) \xi(t) 
\end{align}
If we re-parametrize back to the original notations of Eq.~(\ref{eq:x-eom}), \textit{i.e.} $k \mapsto 1/g$ and $f \mapsto f/g^2$, we obtain
\begin{align}\label{eq:eom3e-alpha}
 \rmd_t x(t) &= - g^2(x) V'(x) + 2D (1-\alpha) \ g'(x)g(x) + g(x) \xi(t) 
\end{align}
that is the same drifted equation that we had obtained from a Fokker-Planck analysis in Sec.~\ref{subsec:FP}, see Eq.~(\ref{eq:x-eom-drifted}).

\section{The path integral}
\label{app}
{Following a route similar to the ones in Refs.~\cite{Janssen79,Langouche79,Langouche81,Janssen92}, 
}
we sketch the construction of the path integral for the
Langevin equation of motion with multiplicative white noise:
\begin{align}
 \mbox{Eq}[x(t),\xi(t)] \equiv \rmd_t x - f(x) - g(x) \xi(t) = 0
 \; . 
\end{align}
In the construction, we use a continuous time  notation with the discretisation subtleties being 
encoded in the choice of the value of the Heaviside theta-function at 
zero, $\Theta(0)=\alpha$. Later, we specify the definition of the path-integral measure.

\paragraph{Path integral construction.}
The explicit calculation of the Jacobian yields
\begin{align}
 \mathcal{J} \equiv  \det_{tt'} \left[ \frac{\delta \mbox{Eq}[x(t),\xi(t)]}{\delta x(t')} \right] = \det_{tt'} \left[ \rmd_t \delta(t-t') + A(x,\xi) \delta(t-t') \right]
\end{align}
with $A(x,\xi) \equiv - f'(x) - g'(x) \xi(t)$, $f'(x) = \rmd_x f(x)$ and $g'(x)=\rmd_x g(x)$. 
Note that if $g(x)\neq \mbox{ct}$ the noise appears explicitly in the functional under $\det_{tt'}$.
After some simple algebra, ${\mathcal J}$ can be factorised as
\begin{align}
 \mathcal{J} \equiv
  \det_{tt'} \left[ \rmd_t \delta(t-t') \right] \, 
 \det_{tt'} \left[\delta(t-t') + \Theta(t-t') A(x,\xi) \right] 
 \; , \label{eq:Jac2}
\end{align}
and the first factor can be discarded in the normalisation. 
We can now re-write the second factor with the help of the identity 
$\det (1+C_\xi) = \exp \mbox{Tr} \ln (1 + C_\xi)$ with the causal function $C_\xi(x,t,t') = \Theta(t-t')A(x,\xi)$,
where we highlighted the dependence of $C_\xi$ on the noise by adding a subscript $\xi$ to $C$.
The $\ln (1+C_\xi)$ can now be expanded in Taylor series. Usually, the causal structure of 
$C$ (that is also usually noise-independent)
truncates the series at first order in $C$. However, in this explicitly noise dependent case
one needs to be careful and also keep the quadratic order~\cite{Arnold2000}:
\begin{align}
 \mathcal{J} & 
 \propto  
 \exp \mbox{Tr}_{tt'} \left[  \Theta(t-t') A(x,\xi) - \frac{1}{2} C_\xi^2(x,t,t') \right] 
 \nonumber
 \\
 & = 
  \exp \int {\rm d}t \left[  \Theta(0) A(x,\xi) - \frac{1}{2} C_\xi^2(x,t,t) \right]
\end{align}
where $C_\xi^2(x,t,t') \equiv \int\ud{t''}  \! \Theta(t-t'') A(x(t),\xi(t)) \, \Theta(t''-t') A(x(t''),\xi(t''))$.
Using now $\Theta(0)= \alpha$, and 
simplifying notations such as $\dot x = \rmd_t x$, $g'(x(t)) = g'$, $C_\xi^2(x,t,t') = C_\xi^2$ and 
$\int\ud{t} = \int$, $P[x]$ reads
\begin{align}
 P[x] &\propto  
 \int\uD[\xi] {\cal D}[\hat{x}] \
 \rme^{
  \alpha\int A(x,\xi)  
  -\frac{1}{2} {\rm Tr}_{tt'}C_\xi^2
  - \int \rmi\hat{x} \left[ \dot x - f - g \xi \right]
  -\frac{1}{4D} \int \xi^2
}
\; . 
\end{align}
(To alleviate the notation we do not write here the time-dependence of the functions in the action, as we do in the main text.)
Before performing the integration over $\xi$ that involves
\begin{align}
& \int\uD{[\xi]} \
\rme^{
-\frac{1}{4D} \int \xi^2 
+\int \left( \rmi\hat{x} g - \alpha g' \right) \xi
-\frac12  {\rm Tr}_{tt'} C_\xi^2
}
\; , 
 \label{eq:26}
\end{align}
let us translate the noise by a function of the variables $x$ and $\rmi \hat x$,
$\xi \mapsto \xi + 2 D \left( \rmi\hat{x} g - \alpha g' \right)$, in the functional integral. 
Notice that $\xi \in \mathbb{R}$ but $\rmi\hat{x} \in \rmi\mathbb{R}$. We can restore the original  
integration domain using the analyticity of the exponential that is zero on the 
boundary thanks to the term $-(4D)^{-1} \int \xi^2$.
The functional  integral in~(\ref{eq:26}) transforms into a new path integral 
\begin{align}
& \rme^{-D \int \left( \rmi\hat{x} g - \alpha g' \right)^2}
\int\uD{[\xi]} \
\rme^{
-\frac{1}{4D} \int \xi^2 
-\frac12  {\rm Tr}_{tt'} C^2_{\xi +2 D \left( \rmi\hat{x} g - \alpha g' \right)}
} 
\; . 
\end{align}
Keeping the terms in $C^2_{\xi+2D\left( \rmi\hat{x} g - \alpha g' \right)}$
that are quadratic in the noise and yield a $\delta(t-t')$ contribution 
within the ${\rm Tr}_{tt'}$ under the noise average, 
and using the notation $\langle \dots \rangle = \int {\cal D}[\xi] \ \rme^{-(4D)^{-1} \int \xi^2}\dots$
one has
\begin{align*}
\langle 
& \rme^{
-\frac12  {\rm Tr}_{tt'} C^2_{\xi + 2 D \left( \rmi\hat{x} g - \alpha g' \right)}
}
\rangle  
=
\rme^{
-\frac12 \langle  {\rm Tr}_{tt'} C^2_{\xi + 2 D \left( \rmi\hat{x} g - \alpha g' \right)}\rangle
} 
\nonumber\\
& 
\qquad\qquad
= 
\rme^{-\frac12 \iint\udd{t}{t'}
 \Theta(t-t') \Theta(t'-t) g'(x(t)) g'(x(t'))
 \langle 
 \xi(t) \xi(t')
 \rangle }
 = \rme^{ - D \alpha^2 \int g'^2 }
 \; . 
\end{align*}
Altogether we obtain
$
P[x]\propto \int\uD[\hat{x}] \ \rme^{S[x,\rmi\hat{x}]}
$
with the action
\begin{align}
 S[x,\rmi\hat{x}] & 
 = -
 \int \left[ \rmi\hat{x} ( \dot x - f + 2D\alpha g' g - D  \rmi\hat{x}  g^2  )
+ \alpha f' \right] 
 \label{eq:x-action-1}
\end{align}
(to which we should add the contribution from the initial measure).
{
This action is consistent with the results reported in Ref.~\cite{Lubensky2007} 
who used  a slightly different approach in which the equation of motion was 
reformulated as}
\begin{align}
 \mbox{Eq}[x(t),\xi(t)] \equiv \frac{\rmd_t x - f(x)}{g(x)} - \xi(t) = 0
 \; . 
 \label{eq:x-eom1}
\end{align}
This is convenient because the noise does not appear explicitly in the Jacobian,
although its effect subtly re-appears along the calculation

\paragraph{Path integral measure.}

We work with a symmetric time interval $t\in [-\mathcal{T},\mathcal{T}]$
which is divided in $N$ discrete time intervals, $t_n \equiv -\mathcal{T} +
n \Delta t$ with $n=0, \ldots, N$ and increment $\Delta t \equiv
2\mathcal{T}/N$. The continuous time
limit is performed by sending $N$ to infinity while keeping ${\cal T}$
finite.
We define the path integral over trajectories on the time interval
$[-\mathcal{T},\mathcal{T}]$ as
\begin{align}
\int \mathcal{D}[x]  = \lim\limits_{N\to\infty} \prod_{n=0}^N \int \rmd x_n
\end{align}
and for the auxiliary field
\begin{align}
 \int \mathcal{D}[\hat{x}]  = \lim\limits_{N\to\infty} \prod_{n=1}^N \int
\frac{\rmd \hat{x}_n}{2 \pi}\;.
\end{align}

\vspace{0.5cm}

\noindent
{\rm \bf Acknowledgements.}
We thank F. Rom\'a, J. Lutsko, S.-I. Sasa, M. Itami and R. Chetrite for useful discussions.
We acknowledge financial support from the  NSF grant No. DMR-115181, PICT-2012-0172, PIP CONICET 2012 0931 (Argentina), the ECOS-Sud A14E01 collaboration, the CNRS-CONICET collaboration PICS 506691, and the bi-national collaboration FAPERJ/CONICET 2014 (Brazil-Argentina). We also acknowledge the Brazilian agencies FAPERJ, CNPq and CAPES for partial financial support. DGB, LFC and GSL thank the International Centre for Theoretical Physics, Trieste, Italy, for hospitality. DGB is Senior Associate to the International Centre for Theoretical Physics, Trieste, Italy. LFC is a member of Institut Universitaire de France. ZGA is postdoctoral fellow PNPD-CAPES-UERJ.
\vspace{0.2cm}

\bibliographystyle{phaip}

\bibliography{magn-FDT}

\end{document}